\documentclass[aps,prd,superscriptaddress,nofootinbib,tighten,preprint]{revtex4}
\usepackage[utf8]{inputenc}
\usepackage[english]{babel}
\usepackage{amsmath}
\usepackage{subfigure}
\usepackage{cancel}
\usepackage{amsfonts}
\usepackage{amssymb}
\usepackage{graphicx}
\usepackage{slashed}
\usepackage{xcolor,color}
\usepackage{mathtools}
\usepackage{resizegather}
\newcommand{\be}{\begin{equation}}
\newcommand{\ee}{\end{equation}}
\newcommand{\bea}{\begin{eqnarray}}
\newcommand{\eea}{\end{eqnarray}}

\newcommand{\nn}{\nonumber}

\catcode`@=12

\begin{document}
\preprint{HRI-RECAPP-2025-02}

\title{Multicomponent Dark Matter with Collider Implications}

\author{Laura Covi}
\email{laura.covi@uni-goettingen.de}
\affiliation{Institut f\"{u}r Theoretische Physik, 
Georg-August-Universit\"{a}t G\"{o}ttingen, Friedrich-Hund-Platz 1,
37077 G\"{o}ttingen, Germany}
\author{Shyamashish Dey}
\email{shyamashishdey@hri.res.in}
\affiliation{Regional Centre for Accelerator-based Particle Physics, Harish-Chandra Research Institute,
A CI of Homi Bhabha National Institute,
Chhatnag Road, Jhunsi, Prayagraj 211019, India}
\author{Sarif Khan}
\email{sarifkhan@cau.ac.kr}
\affiliation{Institut f\"{u}r Theoretische Physik, 
Georg-August-Universit\"{a}t G\"{o}ttingen, Friedrich-Hund-Platz 1,
37077 G\"{o}ttingen, Germany}
\affiliation{Department of Physics, Chung-Ang University, Seoul 06974, Korea}
\author{Santosh Kumar Rai}
\email{skrai@hri.res.in}
\affiliation{Regional Centre for Accelerator-based Particle Physics, Harish-Chandra Research Institute,
A CI of Homi Bhabha National Institute,
Chhatnag Road, Jhunsi, Prayagraj 211019, India}

\begin{abstract} 

The present work aims to study an extension of the Standard Model (SM) that addresses the prominent SM shortcomings, 
i.e., can explain the neutrino mass, the dark matter (DM) content, and the matter-antimatter asymmetry of the Universe. 
The model introduces the possibility of a multicomponent DM scenario leading to distinctive signals at colliders.
The SM is extended by a ``dark" $SU(2)_{D}$ gauge symmetry and new fermions and scalar doublets, 
charged only under $SU(2)_{D}$,  that provide candidates for a multicomponent DM.
Previously, we have considered in this model the asymmetric DM scenario, while in the present work,
we explore the symmetric DM case.
We focus on the parameter region where the dark fermion DM annihilates 
dominantly into the additional dark gauge bosons and the ``inert" doublet DM 
annihilates to SM states via the SM Higgs resonance. 
This particular choice of doublet mass ensures that the heavier BSM Higgs always has one 
decay mode open to DM leading to the possibility of detecting such particles at the LHC, 
in the missing energy plus dijet ($\cancel{E}_{T}+2j$) final states. 
We also discuss the prospects for detecting DM through direct and indirect detection experiments
and via Long-Lived-Particle searches.
Finally, we show that, as typical of other WIMP models, for low DM mass, signals can be expected in 
future collider experiments, but for the higher mass range above $500$ GeV we have to rely solely on 
direct detection experiments. Both types of experiments will be essential to fully cover the allowed parameter space.
\end{abstract}
\maketitle
%%%%%%%%%%%%%%%%%%%%%%%%%%%%%%%%%%%%%%%%%%%%%%%%%%%%%%%%%%%%%%%
\section{Introduction}
\label{Intro}

The Standard Model (SM) of particle physics is a highly successful theory in describing the visible world. However, it does not include a particle with the properties of dark matter (DM). Additionally, it fails to explain the matter-antimatter asymmetry of the Universe. The SM also does not provide a mechanism for neutrino masses at the renormalizable level without introducing a right-handed neutrino.
The presence of DM has been confirmed by many evidences that include
the flatness of galaxy rotation curves \cite{Sofue:2000jx}, observation of the Bullet cluster \cite{Clowe:2003tk}, and the precise measurements of the DM relic density by the Planck satellite \cite{Planck:2018vyg}. Additionally, the baryon density of the Universe has been measured with remarkable precision \cite{Fields:2014uja}.
Moreover, experimentally observed neutrino oscillations between different flavours indicate that neutrinos have mass. These observations have also measured the squared mass differences among the neutrino mass eigenstates \cite{Super-Kamiokande:1998kpq, SNO:2002tuh}. 

In this work, we explore an extension of the SM that provides a unified framework to address these challenges. 
The model establishes a connection between neutrino masses and leptogenesis and the nature of dark matter, whether asymmetric or symmetric.
The model has a dark gauge symmetry $SU(2)_{D}$ that is spontaneously broken by an exotic Higgs sector. Additionally, the model also features 
a discrete symmetry $\mathbb{Z}_{3} \times  \mathbb{Z}_{2}$.  The particle content is extended to include two sets of left-handed $SU(2)_{D}$ fermion 
doublets (ensuring that we are safe from the Witten anomaly) along with their right-handed singlet counterparts. Additionally, the model introduces 
two scalar doublets under $SU(2)_{D}$. This constitutes a minimal extension that maintains a consistent dark $SU(2)$ gauge symmetry, with no direct 
mixing between the SM and dark sectors, except through the scalar and neutrino portals.
We have assigned the $\mathbb{Z}_{3}$ charges to prevent mixing between the fermion doublets, their singlet counterparts, and the SM fermions.
Among the two additional scalar doublets, one of them acquires a vacuum expectation
value (v.e.v.) and breaks the $SU(2)_{D}$ spontaneously.
Among the four components of this doublet, three become the longitudinal component of the massive dark gauge bosons, while the second doublet
remains as an inert doublet with a vanishing vacuum expectation value.
Moreover, we introduce three right-handed (RH) neutrinos to explain the neutrino mass and baryon asymmetry. 
These additional particles are initially neutral under the SM gauge group but mix with SM particles after symmetry breaking.

In Ref. \cite{Biswas:2018sib}, we explored the production of the lepton number
and DM from the decay of the RH neutrino and found that there is
a large parameter space where the model can satisfy the present
observations with an asymmetric DM component if the
mass of the DM is not too large (below about 100 GeV). 
Since the CP asymmetries in the lepton and DM sectors are driven by the same couplings 
and phases, 
they are naturally of similar order and so are the leptonic and DM asymmetries.

In this work, we focus on the scenario where the symmetric DM component dominates and examine the 
model's scalar sector.
A negligible asymmetry for DM can be achieved very easily by reducing the Yukawa coupling between dark matter 
and RH neutrinos, without affecting the neutrino masses, mixings, and the lepton asymmetry. As a symmetric DM, 
the model offers multiple DM candidates, i.e. the exotic fermions, the extra gauge bosons, and the inert doublet, 
allowing a multi-component DM scenario.
We focus here on the case of only two DM components as mass-degenerate fermionic and scalar components 
setting the extra gauge boson mass between the inert doublet and fermion masses so that the gauge bosons decay 
to inert doublets quickly on cosmological scales. 
This allows us to study the fermion doublets as multiple degrees of freedom of a single doublet state and
easily obtain the total fermionic relic density after freeze-out. Note that two of the four Dirac fermions finally 
decay into the scalar inert doublet and contribute to the scalar DM component. 

In this work, we also explore the prospects for DM detection in different experiments, namely direct detection, 
indirect detection, and collider experiments. At the colliders, the BSM Higgs can be produced only through the 
mixing with the SM Higgs.  The BSM Higgs then decays leading to the possibility of different signals, 
depending on the mass spectrum.
We focus here on the scenario where the exotic Higgs decays dominantly
into the
dark sector with the primary collider signature being missing energy
accompanied
by jets. This scenario is realised if the fermionic Dark Matter
substantially annihilates
into exotic dark gauge bosons, while the light inert scalar doublet lies
within the SM Higgs resonance region.

The rest of the paper is organised as follows: in section II, we review the model and
the bounds that we apply to the DM candidates. In section III, 
we consider in
detail the symmetric DM production via freeze-out for the case of light inert
scalar and not too heavy fermionic DM. In section IV, we discuss the collider
signals of this particular scenario via the production of the heavy BSM Higgs,
while in section V, we consider the signals at MATHUSLA and CMS for the 
long-lived particles in the dark sector. Finally, we conclude in section VI.

%%%%%%%%%%%%%%%%%%%%%%%%%%%%%%%%%%%%%%%%%%%%%%%%%%%%%%%%%%%%%%%
\section{Model}
\label{model}

We consider the model (presented in Ref.~\cite{Biswas:2018sib} for the production of the asymmetric DM) where 
the SM gauge group is extended by an additional $SU(2)_{D}$ and the two discrete symmetry groups,
$\mathbb{Z}_{3} \times \mathbb{Z}_{2}$ and the particle content by two left-handed fermion 
doublets and their RH singlet counterparts, and two complex scalar doublets under $SU(2)_{D}$.
The discrete symmetries are added to avoid mixing among the exotic fermions and between the SM and the dark sector. 
Thanks to these symmetries, it is clear that two different states of the dark sector, with different discrete charges, 
are stable and therefore possible DM candidates.

\def\I{i}
\begin{center}
\begin{table}[h!]
%\begin{center}
\begin{tabular}{||c|c|c||}
%\topline
\hline
\hline
\begin{tabular}{c}
    %\hline
    %\multicolumn{1}{c}{}\\
    %\hline
    Gauge\\
    Group\\ 
    \hline
    SU(3)$_{c}$\\ 
    \hline
    SU(2)$_{L}$\\ 
    \hline
    SU(2)$_D$\\
    \hline
    $(\mathbb{Z}_{3},\mathbb{Z}_{2})$\\     
\end{tabular}
&
\begin{tabular}{c|c|c|c|c|c|c}
    %\hline
    \multicolumn{6}{c}{Fermion Fields}\\
    \hline
    ${\Psi_1}_{L} =(\psi_{1},\psi_{2})_L^{T}$ & ~$\psi_{1\,R}$~ & ~$\psi_{2\,R}$~ & 
${\Psi_2}_{L}=(\psi_{3},\psi_{4})_L^{T}$ & ~$\psi_{3\,R}$~ & ~$\psi_{4\,R}$~ & ~${N_i}$~ \\
    \hline
    $1$&$1$&$1$&$1$&$1$&1&1\\
    \hline
    $1$&$1$&$1$&$1$&$1$&1&1\\
\hline
    $2$&$1$&$1$&$2$&$1$&1&1\\
    \hline
    $(\omega,1)$&$(\omega,1)$&$(\omega,1)$&$(\omega^{2},-1)$ & $(\omega^{2},-1)$&$(\omega^{2},-1)$&$(1,1)$\\
    
\end{tabular}
&
\begin{tabular}{c|c|c}
    %\hline
    \multicolumn{3}{c}{Scalar Fields}\\
    \hline
    $\phi_{h}$&$\phi_D$&$\eta_D$\\
    \hline
    ~~$1$~~&$1$&1\\
\hline
    ~~$2$~~&$1$&1\\
    \hline
    ~~$1$~~&$2$ & $2$\\
    \hline
    ~~$(1,1)$~~&$(1,1)$ & $(\omega,1)$\\
     
\end{tabular}\\
\hline
\hline
\end{tabular}
%\bottomrule
%\hline
%\hline
\caption{The particle content for the present study and their charges under different
gauge groups and discrete symmetries. The $\mathbb{Z}_3$ generator $\omega$
satisfies $w^3 = 1$.} 
\label{tab:tab2}
%\end{center}
\end{table}
\end{center}  

In Table \ref{tab:tab2}, we show the new particles and their charges.
We include here the SM Higgs field as well, as it mixes with the dark scalars and with the
RH neutrino provides a connection between the SM and dark sectors.
The SU(3)$_{c} \times {\rm SU(2)}_{L} \times {\rm SU(2)}_{D} \times {\rm U(1)}_{Y}
\times \mathbb{Z}_{3}  \times \mathbb{Z}_{2} $ invariant Lagrangian
is given by,
\begin{eqnarray}
&& \mathcal{L}  =  \mathcal{L}_{SM} - \frac{1}{2} {\rm Tr}  ( F_{\mu\nu}^D F_D^{\mu\nu} )
+ \left( D^{D}_{\mu} \phi_{D} \right)^{\dagger}
\left( D^{D}_{\mu} \phi_{D} \right) 
+ \left( D^{D}_{\mu} \eta_{D} \right)^{\dagger}
\left( D^{D}_{\mu} \eta_{D} \right) 
- \mathcal{V}(\phi_h,\phi_D,\eta_D)
\nn \\
&&
+ i \,\overline{\Psi_k} \gamma^{\mu} D^{D}_{\mu}\,\Psi_k  
+ i \,\overline{\psi_k}\; \slashed\partial\,\psi_k 
+i\,\overline{{N_{j}}_R}\,\slashed{\partial}\,{N_{j}}_R 
 - M_{j} \overline{{N_j}_R^c} {N_{j}}_R  
-  \left(y_{ij} \bar{L}_i \tilde{\phi_h} {N_j}_R 
 + \bar \alpha_j \overline{{\Psi_1}_L} \eta_D {N_j}_R + {\it h.c.}\right)
 \nn \\
&& 
- \left(\lambda_1 \overline{{\Psi_1}_{L}}\,\tilde{\phi_{D}}\,\psi_{1\,R} 
 + \lambda_2 \overline{{\Psi_1}_{L}}\,\phi_{D}\,\psi_{2\,R}
 + \lambda_3 \overline{{\Psi_2}_{L}}\,\tilde{\phi_{D}}\,\psi_{3\,R} 
+ \lambda_4 \overline{{\Psi_2}_{L}}\,\phi_{D}\,\psi_{4\,R} + {\it h.c.}\right) \,,
\nn
\label{Eq:lag}
\end{eqnarray}
where $\mathcal{L}_{SM}$ represents the Lagrangian for the SM sector, while the remaining terms correspond to the kinetic and Yukawa interactions of the BSM particles, including the standard Yukawa coupling between leptons and the RH neutrino. $D^{D}_{\mu}$
in the Lagrangian is the co-variant derivative  associated with the $SU(2)_{D}$ 
gauge symmetry.  
The potential for the scalar fields, including the SM Higgs, is given by 
% $V(\phi_{h}, \phi_{D},\eta_{D})$
\begin{eqnarray}
\mathcal{V}(\phi_h,\phi_D,\eta_D) &=& -\mu_h^2 (\phi_h^{\dagger} \phi_h)
+ \lambda_h (\phi_h^{\dagger} \phi_h)^2 -\mu_D^2 (\phi_D^{\dagger} \phi_D)
+ \lambda_D (\phi_D^{\dagger} \phi_D)^2 \nn \\
&& + \mu_{\eta}^2 (\eta_D^{\dagger} \eta_D)
+ \lambda_{\eta} (\eta_D^{\dagger} \eta_D)^2
+ \lambda_{hD} (\phi_h^{\dagger} \phi_h)(\phi_D^{\dagger} \phi_D)
+ \lambda_{h\eta} (\phi_h^{\dagger} \phi_h)(\eta_D^{\dagger} \eta_D) \nn \\
&& + \lambda_{D1} (\phi_D^{\dagger} \phi_D)(\eta_D^{\dagger} \eta_D)
+ \lambda_{D2} (\phi_{D}^{\dagger} \eta_D) (\eta_D^{\dagger} \phi_{D})\,.
\label{Eq:vscalar}
\end{eqnarray}

We neglect here the possible non-renormalisable coupling between the exotic scalars, cubic in the inert doublet, 
which can effectively reduce its number density and avoid the possibility of scalar DM.  
Among the two additional scalar doublets, $ \phi_D $ can acquire a v.e.v. 
spontaneously along with the SM Higgs doublet, 
%due to the negative mass parameters, so they 
and are represented in the unitary gauge as, 
\begin{eqnarray}
\phi_h = \left(\begin{array}{c}
0 \\
\frac{v+h}{\sqrt{2}}  
\end{array}\right),\,\,\,
\phi_D = \left(\begin{array}{c}
0 \\
\frac{v_{D} + H}{\sqrt{2}}  
\end{array}\right),\,\,\,
\eta_D = \left(\begin{array}{c}
\eta_{1} \\
\eta_{2}  
\end{array}\right) \,\,.
\label{doublet-vev}
\end{eqnarray}

After the spontaneous symmetry breaking, we can write down the 
scalar mass matrix in the $(h\,\,\,H)$ basis as,

\begin{eqnarray}
M^2_{scalar} = \left(\begin{array}{cc}
2 \lambda_{h} v^{2} & \lambda_{hD} v\, v_{D}\\
\lambda_{hD} v\, v_{D} & 2 \lambda_{D} v_{D}^{2}  
\end{array}\right).\,\,\,
\label{mass-square}
\end{eqnarray}
We can diagonalise the above mass matrix and define the mass basis in terms 
of the flavour basis as,
\begin{eqnarray}
h_1 &=& h \cos \alpha - H \sin \alpha\,, \nn \\
h_2 &=& h \sin \alpha + H \cos \alpha\,,
\end{eqnarray}
where the mixing angle and mass eigenvalues can be expressed as,
\begin{eqnarray}
\tan 2\alpha &=& \frac{\lambda_{hD} v v_{D}}
{\lambda_{D} v_{D}^2 - \lambda_{h} v^2}\, , \nonumber \\
M_{h_{1}}^2 &=& \lambda_{h} v^2 + \lambda_{D} v_{D}^2 -
\sqrt{(\lambda_{D} v_{D}^2 - \lambda_{h} v^2)^2 + (\lambda_{hD}\, v\, v_{D})}\,, \nn \\
M_{h_{2}}^2 &=& \lambda_{h} v^2 + \lambda_{D} v_{D}^2 +
\sqrt{(\lambda_{D} v_{D}^2 - \lambda_{h} v^2)^2 + (\lambda_{hD}\, v\, v_{D})}\,.
\label{Eq:scalarmass}
\end{eqnarray}
From the above expressions, one can write down the quartic couplings in terms of the mixing angle and mass eigenvalues as,
\begin{eqnarray}
\lambda_{D} &=& \frac{M_{h_2}^2 + M_{h_1}^2 + (M_{h_2}^2 - M_{h_1}^2)\cos 2\alpha}
{4 v_{D}^2}\,, \nn \\
\lambda_{h} &=& \frac{M_{h_2}^2 + M_{h_1}^2 - (M_{h_2}^2 - M_{h_1}^2)\cos 2\alpha}
{4 v_{D}^2}\,, \nn \\
\lambda_{hD} &=& \frac{(M_{h_2}^2 - M_{h_1}^2)\sin 2\alpha}{2 v v_{D}}\,, \nn \\
\mu_h^{2} &=& \lambda_h {v^2} + \lambda_{hD} \frac{v_{D}^2}{2}\,, \nn \\
\mu_D^{2} &=& \lambda_D {v_D^2} + \lambda_{hD} \frac{v^2}{2}\,.
\end{eqnarray}

Both inert scalars $ \eta_{1,2} $ obtain a mass shift from symmetry breaking as
\begin{eqnarray}
M^2_{\eta_1}  &=& \mu_{\eta}^2 + \frac{\lambda_{h\eta}}{2}  v^2 + \frac{\lambda_{D1}}{2} v_{D}^2 \,, \nn \\
M^2_{\eta_2}  &=& \mu_{\eta}^2 + \frac{\lambda_{h\eta}}{2}  v^2 + \frac{\lambda_{D1}}{2} v_{D}^2 
+ \frac{\lambda_{D2}}{2} v_{D}^2 .
\end{eqnarray}
The mass difference between the two doublet states is determined by the coupling \( \lambda_{D2} \), 
which we will consider negligible in our study. Thus, the two states remain mass degenerate and
can be treated as a simple $ SU(2)_D $ doublet. We will denote the common $ \eta $ scalar mass with 
$ M_\eta $ or $ M_{\eta_1} $ equivalently.

Moreover, when the $SU(2)_{D}$ dark scalar acquires a v.e.v., the fermions
combine into four Dirac fermions defined as 
$\psi^{\prime} = \left( \psi_{1\,L}\,\,\,\psi_{1\,R} \right)$, 
$\chi^{\prime} = \left( \psi_{2\,L}\,\,\,\psi_{2\,R} \right)$,
$\psi = \left( \psi_{3\,L}\,\,\,\psi_{3\,R} \right)$, 
$\chi = \left( \psi_{4\,L}\,\,\,\psi_{4\,R} \right)$, and their mass
terms are given as,
\begin{eqnarray}
M_{\psi^{\prime}} = \frac{\lambda_{1} v_{D}}{\sqrt{2}}\,,
M_{\chi^{\prime}} = \frac{\lambda_{2} v_{D}}{\sqrt{2}}\,,
M_{\psi} = \frac{\lambda_{3} v_{D}}{\sqrt{2}}\,\,\,{\rm and}\,\,\,
M_{\chi} = \frac{\lambda_{4} v_{D}}{\sqrt{2}}\,\,.
\label{dm-mass}
\end{eqnarray}
Here we assume that all the fermions are mass degenerate, corresponding to equal $\lambda_i $ 
and since they have the same coupling with the gauge bosons, they behave as multiple copies of a
single fermion doublet with respect to the gauge symmetry. Indeed the unprimed/primed
fermions contain the two $SU(2)_D$ doublets as LH components, similar to up and down quarks in the SM,
and so the gauge interactions can change $ \psi (\psi^\prime) $ into $ \chi  (\chi^\prime)$.
Note that the $ \psi',\chi' $ fermions do also couple to the RH neutrinos contrary to $ \psi, \chi $, and 
via that Yukawa coupling an asymmetric DM component can be generated from the RH neutrino decay. 
However, we take here those couplings $ \bar \alpha_j $ small enough to render negligible 
both the asymmetric population and the coannihilations between the $ \psi', \chi' $ and $ \eta $ states.
Yet the couplings are large enough to ensure that the primed fermions decay into the $\eta $ scalars 
and neutrinos well before nucleosynthesis, providing an additional scalar DM component.

After symmetry breaking, the three additional gauge bosons $W^{i\;\mu}_{D}$ ($i = 1, 2, 3$) 
acquire the same mass $M_{W_D} = \frac{g_{D}\, v_{D}}{2}$ as well, but their interactions
with the DM particles are still determined by the covariant derivatives and independent
of the v.e.v.. The trilinear vertices with the fermionic and scalar doublets allow for the decay
of the gauge bosons into pairs of fermions or $ \eta \bar \eta $ particles, depending on the
mass scales. 
Via those couplings, the $ W_D $ also have a decay channel into two light neutrinos,
but suppressed both by a loop factor and the small neutrino mixing angle between 
RH and LH neutrinos.

\subsection{Dark Matter candidates and bounds}

In this model, we have multiple stable candidates depending on the mass spectrum of the dark particles. 
As mentioned earlier, we consider the masses of the fermions $\psi$, $\chi$, $\psi^{\prime}$ and
 $\chi^{\prime}$ to be the same, and the masses of the inert scalars $\eta_{1}$ and $\eta_{2}$ to 
 be equal as well. \footnote{If we take different masses for the dark sector particles, we do not 
 expect any qualitative changes in our phenomenology apart from 
 $\mathcal{O}(1)$ changes in the relic
 density. Indeed if some of the states are much heavier than the others, they decouple earlier and 
 decay into the lighter states with the emission of a $W_D $ while the lighter state is still 
 in equilibrium. In that case, they do not contribute to the overall relic density. If instead, the decay of the 
 heavier state happens when the lighter state is decoupled, the latter density is increased by an additional 
 population, which can vary depending on the heavier mass. 
 In the intermediate case, which we discuss here, the annihilation of both states is increased by co-annihilations, 
 but the overall degrees of freedom are also increased and since the 
 particles have the same quantum numbers and interactions, the two effects compensate.} 
 
 As long as we do not break the discrete symmetries, we always have at least two stable particles, the lightest particle in the odd parity sector and the lightest $ ( \omega, 1) $ charged particle 
 under $ \mathbb{Z}_{3}  \times  \mathbb{Z}_{2} $.
 Moreover, the dark gauge bosons can also become nearly stable if they are light enough so that their decay channels in the exotic fermions 
 and scalars are kinematically closed. 
 They do not mix with the SM gauge bosons, contrary to what can happen for $U(1) $ extensions, 
 and therefore their decay into SM states arises at one loop level via the Yukawa's $ \bar \alpha_j $ and the
 RH-LH neutrino mixing and is highly suppressed. 
 We will in the following always consider that the gauge bosons can decay fast enough, before BBN, into dark scalars
 and effectively do not contribute to the DM density, since their abundance at freeze-out is always much smaller than 
 the $ \psi ', \chi' $ abundance.

\begin{itemize}
\item If we consider $M_{\eta_{1}} > M_{\psi^{\prime}}$, then the DM candidates 
are the exotic fermions $\psi, \chi, \psi^{\prime}, \chi^{\prime}$, which contain
particles of both sectors.
This mass spectrum was studied in Ref.~\cite{Biswas:2018sib} when the main DM component was an asymmetric population of $\psi^{\prime}, \chi^{\prime}$ particles.
Ref.~\cite{Biswas:2018sib} also focused on the parameter space where the symmetric DM components from $\psi, \chi, \psi^\prime, \chi^\prime $ were subdominant as the mass of the fermions was light and the $SU(2)_D$ interaction strong enough to reduce the abundance.

\item On the other hand, if we consider $M_{\psi^{\prime}},  M_{\chi^{\prime}} > M_{\eta_{1,2}}$, 
then the stable DM candidates are $\psi, \chi, \eta_{1,2}$. In the present work, we consider this particle spectrum where the DM arises from the WIMP and SuperWIMP mechanisms as a symmetric DM population. This scenario is realised when the Yukawa coupling $ \bar\alpha_{j} $ of the dark fermions with the RH neutrinos is small and the branching fraction of RH neutrinos in dark fermions and scalars is suppressed.
\end{itemize}

Note that in both cases, the only way to access the dark sector at colliders is through mixing within the scalar sector.  
Scalars produced via their interactions with SM states can then decay partially into the dark sector particles to give a 
missing energy signal. The decay rate for the exotic scalars depends on the mixing strength, and depending on the 
mass range considered for $h_{2}$, we can have detection prospects at LHC with 14 TeV centre-of-mass (COM) energy, 
via the missing energy channel. 
Additionally, discovery prospects exist via displaced vertices at LHC and at
the proposed MATHUSLA experiment \cite{Curtin:2018mvb} or the ongoing FASER facility \cite{FASER:2022hcn}, 
where the long-lived $\psi^{\prime}, \chi^\prime $ fermions could decay into scalars and neutrinos.

%%%%%%%%%%%%%%%%%%%%%%%%%%%%%%%%%%%%%%%%%%%%%%%%%%%%%%%%%%%%%%%

In our current investigation, we consider meticulously all 
possible constraints on DM. Our assessment encompasses 
collider, direct detection, and indirect detection bounds. 
Below is a brief overview of these constraints:

\begin{itemize}

\item {\bf Collider bounds:}
 
We focus on the detection of DM at the 14 TeV run of the LHC via the Higgs sector. Indeed thanks to the mixing, the BSM Higgs $ h_2 $
can be produced as the SM Higgs is produced and then can decay mainly to the dark sector particles. 
The present LHC constraints on the model reduce to those on an extended Higgs sector and pure scalar portal Higgs models, 
since the dark sector mainly couples to SM fields via the Higgs field. 
We apply all relevant bounds on the Higgs sector that arise from existing LHC analysis of BSM scalars and constrain the parameter 
space by using the HiggsBound package \cite{Bechtle:2015pma}.
Moreover, if the scalar DM mass is less than half of the SM Higgs mass then we also have bounds from the invisible SM Higgs decay. 
In this context, we have used the HiggsSignal package \cite{Bechtle:2013xfa} which also takes care of the  Higgs signal strength.
We will discuss in the later sections the detection prospects of the DM at the collider with a $3\sigma$ and $5\,\sigma$ statistical significance 
over the background for the high luminosity LHC phase.

\item {\bf Direct Detection:}

The DM particles, both fermionic and scalar, interact with the SM via Higgs-mediated processes.
Therefore, both the DM candidates are in principle visible in the direct detection experiments through the scalar-mediated elastic 
scatterings shown in the left panel (LP)  of Fig. \ref{dd-id}. 
\begin{figure}[ht]
\centering
\includegraphics[angle=0,height=4.5cm,width=8.5cm]{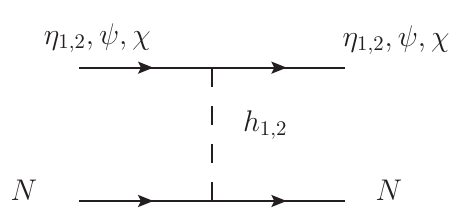}
\includegraphics[angle=0,height=4.5cm,width=8.5cm]{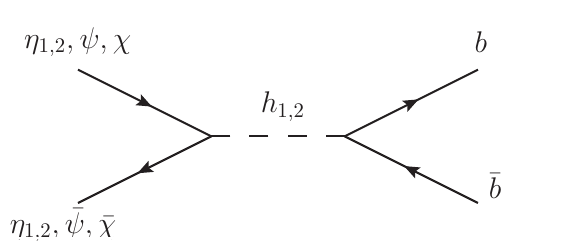}
\caption{LP: Diagram contributing to DM Direct Detection; RP: Diagram giving the
indirect detection signal. Both are mediated by the two Higgses.} 
\label{dd-id}
\end{figure}

The spin-independent direct detection (SIDD) elastic scattering cross-section for the process 
$\eta N \rightarrow \eta N $ ($N$ is nucleon) can be expressed as,
\begin{eqnarray}
\sigma_{SI, scalar} = \frac{\mu^2_{s}}{4 \pi}
\biggl[ \frac{M_{N} f_{N} \cos\alpha}{M_{\eta } v} 
\left(\frac{g_{h_{1} \eta \eta}}{M^2_{h_{1}}} 
+ \frac{ g_{h_{2} \eta \eta} \; \tan\alpha}{M^2_{h_{2}}} 
\right) \biggr]^{2}
\label{SIDD-psi-expression}
\end{eqnarray}
where $\eta = \eta_{1,2}$, $\mu_s = \frac{M_{\eta_{1,2}} M_{N}}{M_{N} + M_{\eta_{1,2}}}$,
$f_{N}$ is the nucleon Higgs coupling, $M_{N}$ is the nucleon mass
and the vertex factors 
$g_{h_{i} \eta_{j} \eta_{j}}\,\, (i,j = 1,2)$ are given as follows,
\begin{eqnarray}
&& g_{h_{1} \eta_{j} \eta_{j}} = -  \lambda_{h\eta} v \cos\alpha +
\left( \lambda_{D1} + \delta_{2j} \lambda_{D2} \right)  v_{D} \sin\alpha\, \nonumber \\
&& g_{h_{2} \eta_{j} \eta_{j}} =    - \lambda_{h\eta} v \sin\alpha
- \left( \lambda_{D1} + \delta_{2j} \lambda_{D2} \right) v_{D} \cos\alpha\,.
\label{h-eta-eta}
\end{eqnarray}
We take here $\lambda_{D2} = 0$, consistent with
degenerate $ \eta_{1,2} $ masses.
The above SIDD cross section holds  for both $\eta_{1,2}$ DM and
after using the explicit expressions, we 
can write down the SIDD cross-section in the following form,
 \begin{eqnarray}
\sigma_{SI, scalar} = \frac{\mu^2_{s}}{4 \pi}
\biggl[ \frac{M_{N} f_{N} }{M_{\eta} v} 
\left( 
 \lambda_{h\eta} v \left(\frac{\cos^{2}\alpha}{M^2_{h_1}} + \frac{\sin^{2}\alpha}{M^2_{h_2}} \right)
-  \lambda_{D1} v_{D} \sin\alpha \cos\alpha  \left(\frac{1}{M^2_{h_1}} - \frac{1}{M^2_{h_2}} \right) 
  \right) \biggr]^{2} \nonumber \\
  \label{SIDD-eta-expression}
\end{eqnarray}
From the above expression, we can see that for very particular values of 
$\lambda_{D1}$, $\alpha$, $M_{h_{2}}$, and $\lambda_{h\eta}$, a
cancellation among the two terms can happen, so that there is in general 
no lower bound on the direct detection cross-section. 
However, the full cancellation is very fine-tuned and does not happen so easily
for parameters that also provide the correct DM density. In general for small mixing
angle, as required by the present Higgs data, the first term related to the exchange of
a SM Higgs gives the dominant contribution and the direct detection cross-section is 
very similar to other Higgs portal models.

Similarly, we can also estimate the SIDD scattering cross-section for the 
fermionic DM candidate as,
\begin{eqnarray}
\sigma_{SI,fermion} = \frac{\mu^2_{f}}{\pi} \biggl[ \frac{M_{N} f_{N} M_{\psi} \sin 2\alpha}
{2\, v\, v_{D} M^2_{h_{1}}} \left( 1 - \frac{M^2_{h_1}}{M^2_{h_2}} \right) \biggr]^{2}
\label{SIDD-psi}
\end{eqnarray}  
where $\mu_f = \frac{M_{\psi,\chi} M_{N}}{M_{N} + M_{\psi,\chi}}$ and the other
quantities are defined earlier.  In this case, the cancellation happens only for degenerate
Higgs masses, but the cross-section is always suppressed by $ \sin 2\alpha $, which is small 
to agree with the SM-like Higgs measurements.

When we compare the direct detection cross-section with different experimental bounds we 
rescale the SIDD cross-section by its present DM fraction as, $\frac{\Omega_{\eta\,( \psi, \chi )}}{\Omega_{tot}} 
\sigma_{SI, scalar(fermion)}$ where $\Omega_{\eta} = \Omega_{\eta_{1} + \eta_{2}}$.
In Fig. \ref{scatter-plot-3} and 
Fig. \ref{scatter-plot-4}, we show the SIDD cross-section for the fermion
and scalar DM candidates, respectively, in the context of the current LUX-ZEPLIN data.

\item {\bf Indirect Detection:}

Both DM candidates can also be detected through indirect detection
experiments by DM annihilating to SM final state particles. As the mediator
is again one of the Higgs scalars, the main DM annihilation is to the $b \bar{b}$ 
final  state for both $\eta \bar \eta $ and $ \psi \bar{\psi}  (\chi \bar{\chi}) $.
The flux of photons or other final state particles is proportional to the cross-section
and to the square of the local DM matter density.
 In our model, since neither of the components amounts to the total DM density, 
 we weight each channel by the relative DM fraction $f^2_{A} $ 
where $ f_{A} =  \frac{\Omega_{A}}{\Omega_{tot}}$. 
The indirect detection processes are shown in the right panel (RP) of Fig. \ref{dd-id}.
In the non-relativistic limit, the indirect detection cross-section 
for the scalar initial state $\eta \bar \eta  \rightarrow b \bar b$ can be expressed as,
\begin{eqnarray}
\left( \sigma v \right)_{\eta  \bar \eta  \rightarrow b \bar{b}} & \simeq &
\frac{n_{c}}{4\pi}  \left( 1- \frac{m^2_{b}}{M^2_\eta} \right)^{3/2}
\biggl[ \frac{g^2_{\eta\eta h_{1}} g^2_{h_{1}bb} }{M^4_{h_{1}} 
\left((1-\xi_{1})^2 + \epsilon_{1}^{2} \right) } 
+ \frac{g^2_{\eta\eta h_{2}} g^2_{h_{2}bb} }{M^4_{h_{2}} 
\left((1-\xi_{2})^2 + \epsilon_{2}^{2} \right) } \nonumber \\
& & + \frac{ 2 g_{\eta\eta h_1}\; g_{h_1 bb}\; g_{\eta\eta h_2}\; g_{h_2bb} \;
(1-\xi_1) (1-\xi_2)}{ M^2_{h_1} M^2_{h_2} 
 \left((1-\xi_{1})^2 + \epsilon_{1}^{2} \right) \left((1-\xi_{2})^2 + \epsilon_{2}^{2} \right) } 
% % \biggl( 16 M^4_{\eta} + % (1 + \epsilon_{1} \epsilon_{2})
 % % M^2_{h_{1}} M^2_{h_{2}} - 4 M^2_{\eta} (M^2_{h_{1}} + M^2_{h_{2}}) \biggr)
  \biggr]\,, 
\label{etaetabb}
\end{eqnarray}
where $\xi_{1,2} = \frac{4 M^2_{\eta}}{M^2_{h_{1,2}}}$, 
$\epsilon_{1,2} = \frac{\Gamma_{h_{1,2}}}{M_{h_{1,2}}}$,
$ g_{\eta \eta h_{1,2}} $ have been defined earlier and
\begin{eqnarray}
g_{h_{1}bb} &=& \frac{M_{b} \cos\alpha  }{v}\,,\,\,
g_{h_{2}bb} =  \frac{ M_{b} \sin\alpha }{v}\; .
\end{eqnarray}
 
In the same way, $\psi \bar \psi \rightarrow b \bar b$ annihilation cross section
times velocity in the non-relativistic limit can be expressed as,
\begin{eqnarray}
\left( \sigma v \right)_{\psi \bar \psi \rightarrow b \bar{b}} &\simeq & 
\frac{n_{c} }{8 \pi} M^2_{\psi} v^2_{rel} \left( 1- \frac{m^2_{b}}{M^2_{\psi}} \right)^{3/2} 
\biggl[ \frac{g^2_{\psi\psi h_{1}} g^2_{h_{1}bb} }{M^4_{h_{1}} 
\left((1-\xi^{\prime}_{1})^2 + \epsilon^{\prime\,2}_{1} \right) } 
+ \frac{g^2_{\psi\psi h_{2}} g^2_{h_{2}bb} }{M^4_{h_{2}} 
\left((1-\xi^{\prime}_{2})^2 + \epsilon^{\prime\,2}_{2} \right) } \nonumber \\
&& + \frac{ 2 g_{\psi\psi h_1}\; g_{h_1 bb}\; g_{\psi\psi h_2}\; g_{h_2 bb}\; (1-\xi_1^\prime) (1-\xi_2^\prime) }{ M^2_{h_1} M^2_{h_2} 
\left((1-\xi^{\prime}_{1})^2 + \epsilon^{\prime\,2}_{1} \right) \left((1-\xi^{\prime}_{2})^2 + \epsilon^{\prime\,2}_{2} \right) } 
  \biggr] 
  \label{psipsibb}
\end{eqnarray} 
where $\xi^{\prime}_{1,2} = \frac{4 M^2_{\psi}}{M^2_{h_{1,2}}}$, 
$\epsilon^{\prime}_{1,2} = \frac{\Gamma_{h_{1,2}}}{M_{h_{1,2}}}$, $g_{\psi\psi h_{1}} = - \frac{ \sin\alpha M_{\psi}}{v_{d}}$ 
 and $g_{\psi\psi h_{2}} = \frac{\cos\alpha M_{\psi}}{v_{d}}$\,.
As in the case of fermionic portal models, the cross-section is velocity suppressed away from the resonance, and therefore typically smaller than the one for the scalar component. 
 In the LP and RP of Fig. \ref{scatter-plot-5}, we show the indirect detection cross section for the fermion and scalar DM, respectively, 
 and the associated bound from the Fermi-LAT data. 
 
\end{itemize}

\section{Dark matter relic density}

In Ref. \cite{Biswas:2018sib}, we studied in detail the production of asymmetric dark matter components 
from the RH neutrino decay into the fermions $ \psi^\prime, \chi^\prime $ with masses up to hundreds of GeV.
In that case, the symmetric fermionic component was negligible thanks to the relatively strong interactions with the dark 
$ SU(2) $ gauge bosons, while in the case of the scalar component, it could be erased either by resonant annihilation 
or by the presence of (non-renormalizable) cubic interactions for $\eta $.
With a different choice of parameters, in particular, a smaller Yukawa coupling $ \bar \alpha_j $, 
we can suppress the asymmetric part and make the symmetric DM component dominant and in agreement with 
the cosmological measurement for larger fermion masses.

The model allows for a multi-component DM scenario depending on the choice of
the masses. Here we consider $M_{\psi} = M_{\chi} > M_{\eta_1} = M_{\eta_2} = M_\eta $
and $M_{W_D} = 155$ GeV allowing us to reduce the problem to only two Boltzmann 
equations associated with the doublets $\psi, \chi $ and $\eta$. For the other fermionic dark states, 
the fermions $ \psi^\prime, \chi^\prime $ will have the same relic density as the unprimed
fermions, but will decay at late times into the inert scalars and neutrinos.

We use the code MicrOMEGAs \cite{Belanger:2001fz} to compute the relic densities and we have implemented 
the present model in FeynRules \cite{Alloul:2013bka} to generate the CalcHEP files for the cross-sections \cite{Belyaev:2012qa}. 
The same model file is also used for generating the UFO file to study the collider phenomenology using 
MadGraph \cite{Alwall:2011uj}. 
The abundance of the DM components is determined by the 
following coupled Boltzmann equations,
\begin{eqnarray}
\frac{d Y_{\eta}}{dz} &=& -\frac{2 \pi^2}{45} \frac{M_{pl} M_{sc} 
\sqrt{g_{*}(z)}}{1.66 z^2} \left[ \sum_{\substack{a,b \in eq}}\langle \sigma v 
\rangle_{\bar \eta \eta  \rightarrow
a b} \left( Y^{2}_{\eta} - (Y^{eq}_{\eta})^2 \right) + \!\!\!\!\!
 \sum_{\substack{q,q' \in BSM}}  \!\!
 \langle \sigma v \rangle_{\eta \eta'  \rightarrow q q'} 
\left( Y_{\eta}^2 - Y_q^2 \right)\right]
\nonumber \\
\frac{d Y_{p}}{dz} &=& -\frac{2 \pi^2}{45} \frac{M_{pl} M_{sc} 
\sqrt{g_{*}(z)}}{1.66 z^2}  \left[ \sum_{\substack{a,b \in eq}}\langle \sigma v 
\rangle_{\bar pp' \rightarrow
a b} \left( Y^2_{p} - (Y^{eq}_{p})^2 \right) + \!\!\!\!\!\!
\sum_{\substack{q, q' \in BSM}} \!\!
\langle \sigma v \rangle_{\bar pp' \rightarrow q q' } 
\left( Y^2_{p} - Y_q^2 \right) \right] . \nonumber\\
\end{eqnarray}
where $ q, q' $ denote any other dark state and $ p, p' = \psi, \chi \; \mbox{or}\; \psi^\prime, \chi^\prime $.
Note that the first terms in the Boltzmann equations sum over all processes with final particles in equilibrium
including coannihilations between members of the same type of multiplets, e.g. $ \eta_{1} $ and $ \eta_{2} $ 
as well as $ \psi (\psi^\prime) $ and $\chi (\chi^\prime) $ and annihilations into the dark gauge bosons, which 
remain in thermal equilibrium until very late and therefore do not need to be considered in a third equation.
The second terms instead take into account conversion processes between the even and odd dark 
sectors via the dark gauge bosons,  e.g. $ \bar \eta_1 \eta_2 \rightarrow \bar \psi \chi $.
The latter terms couple the two equations and can transfer the densities from the scalar to the 
fermion particles or vice-versa.
Note, however, that due to the different discrete charges, there are no co-annihilations between the 
unprimed and primed fermions. For the states that interact with the RH neutrino, i.e. $ \psi',\chi' $ and $ \eta $, 
there is in principle the co-annihilation between a fermion and a scalar, e.g.
 $ \psi' \bar \eta \rightarrow N_R \rightarrow  \nu_L h/Z  $, but it is governed by the couplings $ \bar \alpha_{j} $ 
 that we assume here are subdominant and therefore this co-annihilation is suppressed compared to the 
 other channels. Note though that the coupling $ \bar \alpha_{j} $, even if tiny, allows for the decay of the 
 $ \psi^\prime, \chi^\prime $ fermions into neutrinos and the scalar, leading to additional production 
 of inert scalars at late times via the SuperWIMP mechanism.

The symmetric DM density can then be obtained from the abundance of the single states using the following relation,
\begin{eqnarray}
\Omega^{Sym}_{DM} h^2 = 2.755 \times 10^{8}\sum_{X=\psi,\chi,\eta} \left(\frac{M_{X}}{\rm GeV} \right) Y_{X} \,.
\label{relic-density-formula}
\end{eqnarray}
Finally, the total DM relic density that includes the symmetric and 
anti-symmetric part and late production will be
\begin{eqnarray}
\Omega_{DM}h^{2} =  \Omega^{Sym}_{DM} h^2 + \Omega^{SW}_{DM} h^2 + \Omega^{Asym}_{DM} h^2  \,\,\,
\text{where we can assume}\,\, \Omega^{Asym}_{DM} h^2 \sim 0\,.
\end{eqnarray}
In the present work, we have considered the total DM relic density bound in the following range,
\begin{eqnarray}
10^{-3} \leq \Omega_{DM}h^{2} \leq 0.12\,.
\label{DM-relic-density}
\end{eqnarray}
This large allowed range of DM relic density helps us to scan the model parameters judiciously without losing any interesting part of the parameter space.

\subsection{Relic density of the single components}

As pointed out earlier, the model contains many candidates
for DM. We show here a few abundance plots for the different particles,
which show us the dependence of the individual relic densities on the model parameters. 
In generating the line plots, we have set the model parameters at the following values, 
unless they are varied in the plot, i.e. 
$g_{D} = 1.0$, $M_{h_{2}} = 600$ GeV,
$M_{\psi}\, (= M_{\chi}) = M_{\psi^{\prime}}\,(M_{\chi^{\prime}})  
= 250$ GeV, $M_{\eta_{1}} = M_{\eta_{2}} = 62 $ GeV, 
$ M_{W_D} = 155 $ GeV, $\sin\alpha = 0.1$,
$\lambda_{h\eta} = \lambda_{D1} = 0.1$ and $\lambda_{D2} = 0$. 
\begin{figure}[h!]
\centering
\includegraphics[angle=0,height=7.5cm,width=8.5cm]{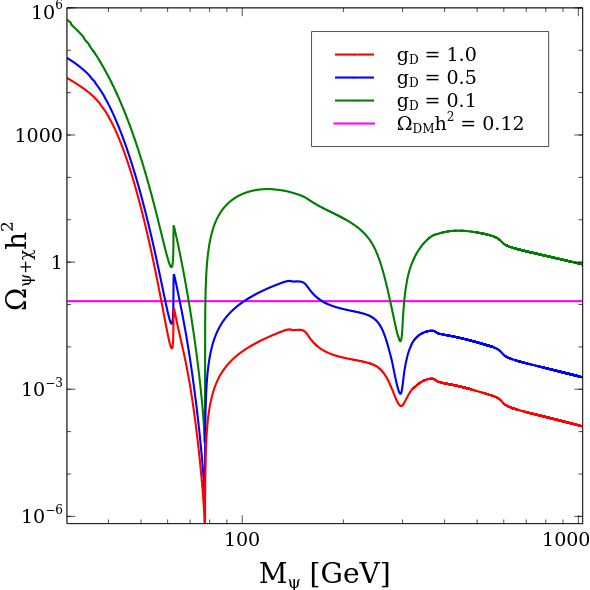}
\includegraphics[angle=0,height=7.5cm,width=8.5cm]{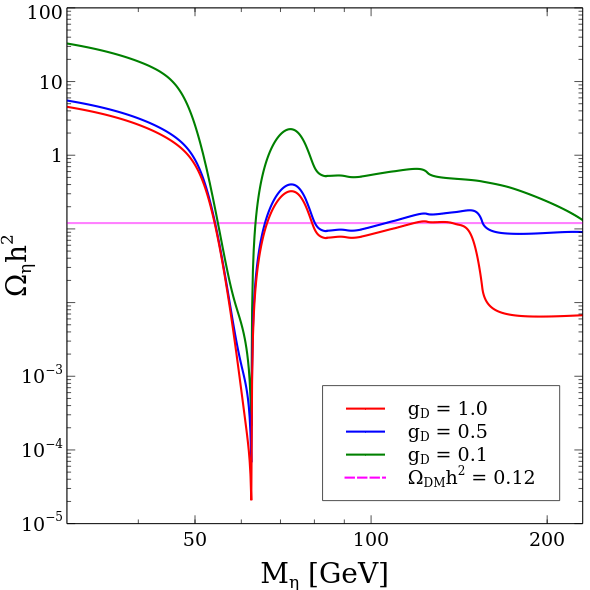}
\caption{In the LP and RP, we show the dependence of the DM relic density
from the fermionic DM mass $M_{\psi}$ and scalar DM mass $M_{\eta}$, respectively.
In the LP and RP, we show the variation of $\psi$ and $\eta$ DM for three different 
values of the gauge coupling $g_{D}$.   
In  generating the plots, we fix the other model parameters
in the LP (RP) as $g_{D} = 1.0$, $M_{W_{i}} = 155$ GeV, $M_{h_{2}} = 600$
GeV, $M_{\eta} = 62$ GeV ($M_{\psi} = 250$ GeV), $\sin\alpha = 0.1$,
$\lambda_{h\eta} = \lambda_{D 1} = 0.1 $. The parameters 
are also kept fixed for the rest of the line plots unless they have been
varied in generating the plots.} 
\label{line-plot-1}
\end{figure}

In the LP and RP of Fig. \ref{line-plot-1}, we show the $\psi,\,\chi$
and $\eta$ DM relic density as a function of their mass, respectively.
In both panels, we give the DM relic density 
for three different values of the $SU(2)_{D}$ gauge coupling $g_{D}$.
In the LP, we have three resonance regions as we increase the mass
of $\psi$ DM, $M_{\psi}$. The first resonance for
$M_{\psi} \sim \frac{M_{h_{1}}}{2}$ represents the SM Higgs resonance
in the channel $\psi \bar \psi \xrightarrow[]{h_{1}} SM\,\,SM $,
the second resonance appears in  
$\psi \bar \psi \xrightarrow[]{W_D}  \eta\,\,\bar \eta$ 
at the $ W_{D} $ mass, while the third resonance region appears due 
to the heavier Higgs $h_2$ resonance in $\psi \bar \psi \xrightarrow[]{h_{2}} SM\,\,SM, BSM\,\,BSM $.
The region between the $W_D$ and $h_2$ resonances is mainly dominated
by the process $ \psi \bar \psi \rightarrow W^{+}W^{-}, ZZ, W_{D}W_{D} $.
In the lower side of the DM mass {\it i.e.} on the left side of the SM Higgs resonance region and below the $ W_{D} , \eta $ mass, the DM has only
annihilation modes into SM particles and since they are suppressed
by the small mixing angle $ \alpha$ and the Higgs propagator, 
the DM density is overproduced. 
As soon as the DM mass is larger than the gauge boson mass, the
annihilation channel into gauge bosons opens up and strongly reduces
the relic density.

In Fig. \ref{line-plot-1}, we also show the DM relic density for 
different values of $g_{D}$ keeping the gauge boson mass fixed and therefore changing the v.e.v. of the dark Higgs doublet $v_{D} = M_{W_{D}}/g_{D}$.
 One important thing to note here is that the dip seen in the
 BSM Higgs resonance is reduced for large $g_{D}$, while 
 the decay width of the dark Higgs increases as we increase the 
 gauge coupling $g_{D}$, making the resonance broader.  
  
In the RP of Fig. \ref{line-plot-1}, we show the $\eta$ DM relic density as a function of its mass $M_{\eta}$. 
We see that the $\eta $ DM relic density varies inversely with the strength of the gauge coupling. This is because 
 $g_{ h_{1} \eta_{1} \eta_{1}  } \propto (v\,\cos\alpha -  M_{W_D} \frac{\sin\alpha}{g_{D}})$ 
 and the value of this coupling is reduced due to a cancellation between the two terms for small $ g_D$.
 We see two dips at $M_{\eta} \sim M_{W^{\pm}}$ and 
 $M_{\eta} \sim M_{W_D} $ due to the opening of the DM annihilation channel
into the SM and dark gauge bosons, but only one resonance for the SM Higgs,
because the heavier Higgs resonance is at larger masses and the
resonant annihilation via dark $W_D$ is kinematically suppressed.
We can see that for $M_{\eta} > \frac{M_{W_{D}}}{2} \sim 74.5$ GeV as well as if we demand 
$M_{\psi, \chi} > \frac{M_{W_{D}}}{2} \sim 74.5$, then $W_{D}$ becomes a metastable DM candidate 
and can potentially contribute to the DM density after annihilating to lighter particles. 
In this mass regime, we can have three-component DM candidates, and we aim to pursue this in the future. 
In the present work, we assume that $ W_D$ always decays before
BBN to $\eta$ or $\psi, \chi$ depending on the kinematics.

\begin{figure}[h!]
\centering
\includegraphics[angle=0,height=7.5cm,width=8.5cm]{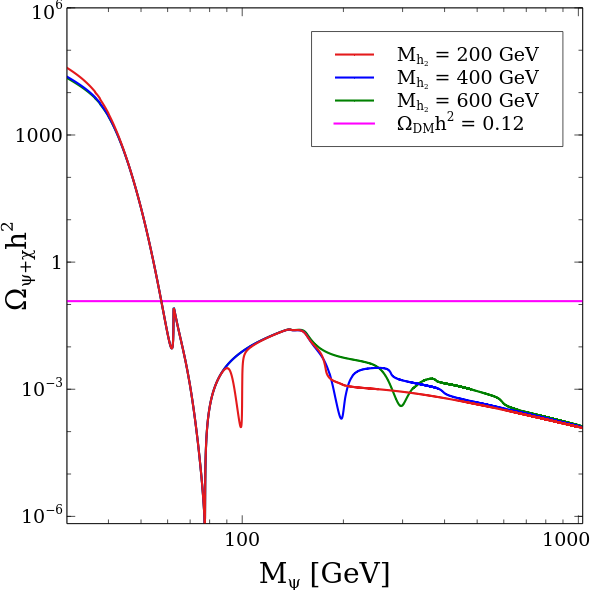}
\includegraphics[angle=0,height=7.5cm,width=8.5cm]{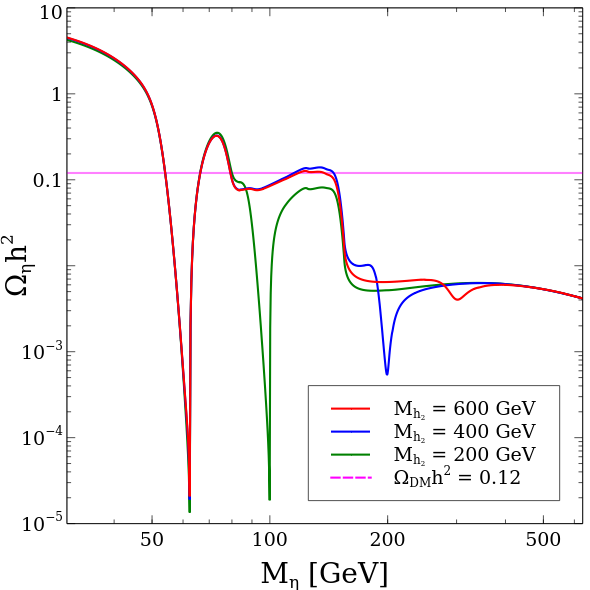}
\caption{The variation in both the plots are same as before. 
In the LP and RP, we have shown variation
with three different values $M_{h_{2}}$.} 
\label{line-plot-2}
\end{figure}

In Fig. \ref{line-plot-2}, we plot the relic densities for three different values of BSM Higgs mass $M_{h_{2}}$.
In the LP (RP), we show the $\psi, \chi \,\, (\eta)$ relic density. In the LP, we still obtain three resonance 
regions as discussed in the previous plot, with the third resonance shifted further to the right 
for larger BSM Higgs mass.
In the RP, we observe new resonances that appear for $ 2 M_\eta = M_{h_2} $.
In particular for $M_{h_{2}} =  200$ GeV, the BSM Higgs decay width is small so the resonance peak is very sharp compared to $M_{h_{2}} = 400, 600$ GeV. 
For the larger values of BSM Higgs mass its decay into the dark gauge
bosons open up and increase the width significantly.

\begin{figure}[h!]
\centering
\includegraphics[angle=0,height=7.5cm,width=8.5cm]{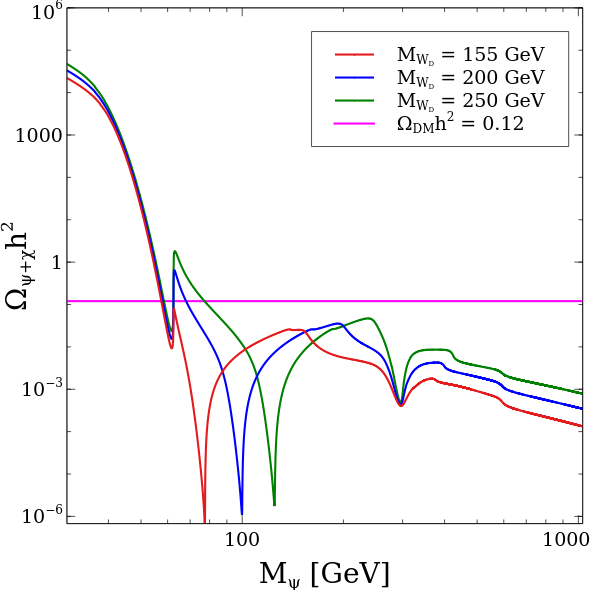}
\includegraphics[angle=0,height=7.5cm,width=8.5cm]{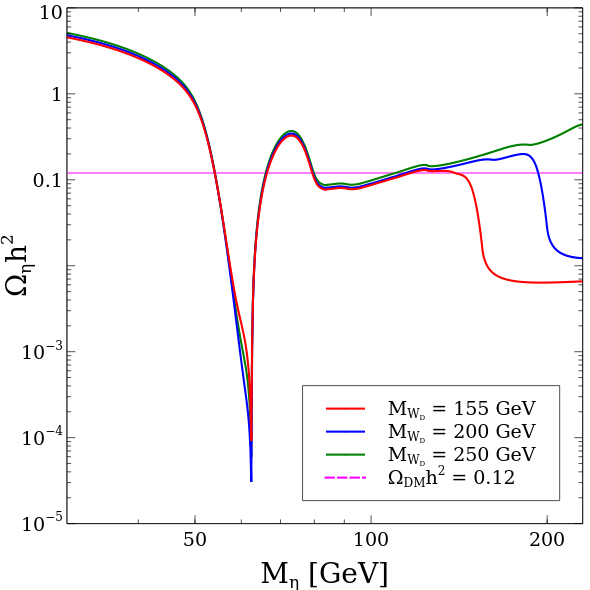}
\caption{In the LP and RP, we have shown the DM relic density variation
w.r.t three different values of additional gauge bosons mass $M_{W_{D}}$.} 
\label{line-plot-3}
\end{figure}

In the LP and RP of Fig. \ref{line-plot-3}, we study the dependence 
of the relic densities on the gauge boson mass $M_{W_D}$. 
In the LP, we can see the three resonance regions and the shift of the 
second resonance with the change of the $M_{W_D}$ mass. 
For very low mass of $\psi,\chi$ DM, we can see a slight change in the $\psi+\chi$ 
density which occurs due to the change of the v.e.v. $v_{D}$. 
Note that the dip of the third resonance does not depend on the mass of the 
dark gauge boson, but only on the gauge coupling.
The broadening in the relic density around the third resonance is due to 
the opening of the DM annihilation to the dark gauge boson, 
$\psi \bar \psi \rightarrow W_D W_D $ and is proportional to the v.e.v. of 
the dark scalar doublet. 
In the RP, we see that the dark gauge boson does not have a 
strong effect on the $\eta$ DM density apart from 
shift of the $\eta \bar \eta \rightarrow W_D W_D $
annihilation mode opening on the right of the plot. 

\begin{figure}[h!]
\centering
\includegraphics[angle=0,height=7.5cm,width=8.5cm]{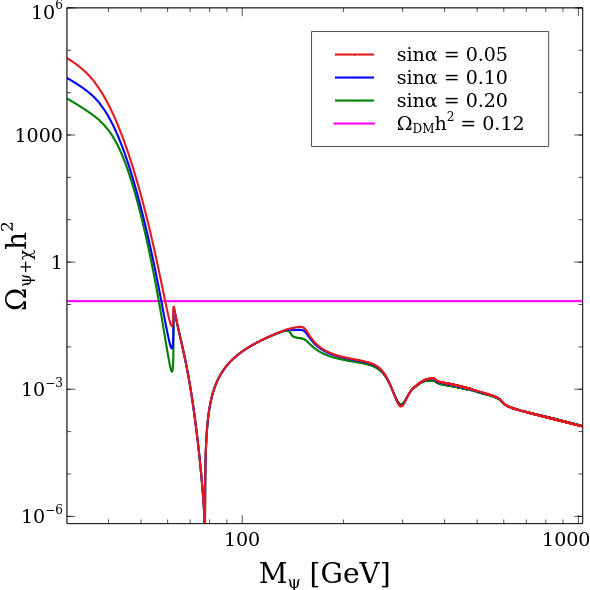}
\includegraphics[angle=0,height=7.5cm,width=8.5cm]{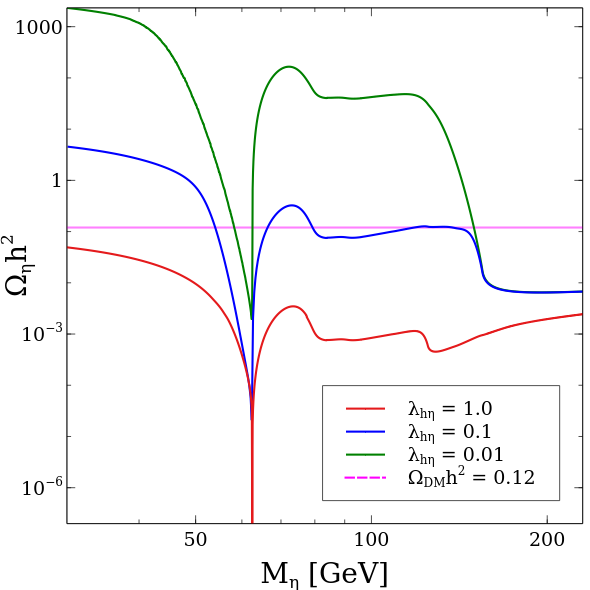}
\caption{LP shows the DM relic density variation for three different
values of Higgses mixing angle $\sin\alpha$ whereas in the RP
we have show the variation of $\eta$ DM density for three different values
$\lambda_{h\eta}$.} 
\label{line-plot-4}
\end{figure}

In the LP of Fig. \ref{line-plot-4}, we plot the DM relic densities
for three different values of the neutral CP-even Higgs mixing angle $\sin\alpha$. 
The $\psi, \chi$ DM annihilation via intermediate Higgs is directly proportional
to the mixing angle $\sin\alpha$, so the first resonance is affected.
The second and third resonances have minimal dependence on the mixing angle and 
are mainly governed by the $SU(2)_{D}$ gauge boson mass and gauge coupling.
In the RP of Fig. \ref{line-plot-4}, we present relic density of $\eta$ for three different values of the portal quartic coupling $\lambda_{h\eta}$. 
This coupling has a significant impact on the scalar relic density, as it scales inversely with $\lambda_{h\eta}^2$.
Moreover, in the region $M_{\eta} > M_{W_D} $, and for smaller values of $\lambda_{h\eta}$ the dominant 
annihilation channel is $\eta \bar \eta \rightarrow W_D W_D $. 
However, for higher values of $\lambda_{h\eta}$ the process $\eta \bar\eta \rightarrow h_{1}h_{1}$ also contributes significantly.

\subsection{Total dark matter density and scatter plots}

We have analysed the response of the relic densities of $\psi, \chi$ and $\eta$ 
w.r.t a few individual model parameters. Now we extend our 
analysis of the entire parameter space and look for correlations among parameters when they are varied simultaneously.
We also highlight the detection prospects at different DM experiments including collider, spaced-based experiments to terrestrial experiments. 

The DM candidates are  $\psi, \chi,\eta$, while $\psi^{\prime}, \chi^{\prime}$ are not stable and decay to $\eta$ at late time. 
The total DM relic density, therefore, is given as
\begin{eqnarray}
\Omega_{DM} h^{2} =  \left( \Omega_{\psi+\chi} + \Omega_{\eta} \right) h^{2}
+ \frac{M_\eta}{M_{\psi^\prime}}  \Omega_{\psi^\prime + \chi^\prime} h^2
\end{eqnarray}
where the last contribution is due to the SuperWIMP production of $ \eta$.
We look at the cosmologically allowed range for the DM relic density
to have possible correlations among the model parameters as follows,
\begin{eqnarray}
10^{-3} \leq \Omega_{DM} h^{2} \leq 0.12\,.
\end{eqnarray}
We vary the model parameters in the following range to generate
scatter plots,
\begin{eqnarray}
&& 130 \leq M_{h_2} \,\,[{\rm GeV}] \leq 1030\,,\,\,
1 \leq \left( M_{\psi,\chi} - M_{\eta_{1}} \right)\,\,[{\rm GeV}] 
\leq 1000\,,40 \leq M_{\eta_{1,2}}\,\,[{\rm GeV}] \leq 80\,,
\nn \\
&& 
10^{-2} \leq g_{D} \leq 1\,,
10^{-3} \leq \sin\alpha \leq 0.5\,,\,\, 10^{-3} \leq \lambda_{h\eta}, \lambda_{D1} \leq 10^{-1}\,,\,\, \lambda_{D2} = 0,\,\,M_{W_{D}} = 155 \,\,{\rm GeV}
\nonumber \\
\label{range-of-parameter}
\end{eqnarray}
To avoid the region where $ W_D $ is metastable, we restrict the study to $M_{W_{D}} > 2 M_{\eta_{1}}$. 

\begin{figure}[h!]
\centering
\includegraphics[angle=0,height=7.5cm,width=8.5cm]{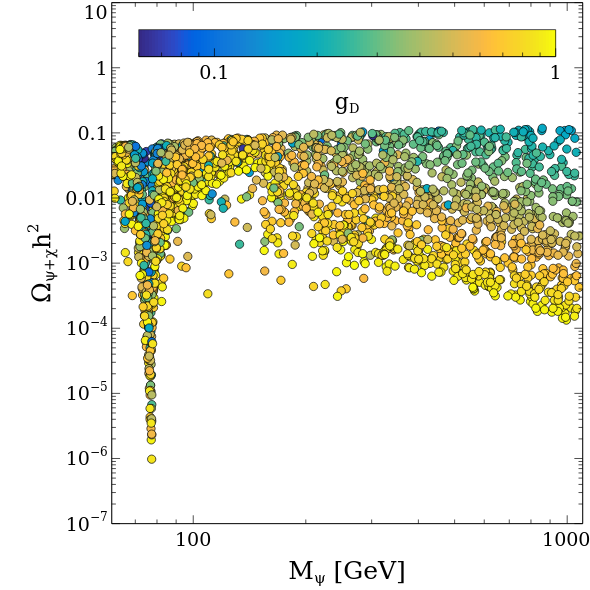}
\includegraphics[angle=0,height=7.5cm,width=8.5cm]{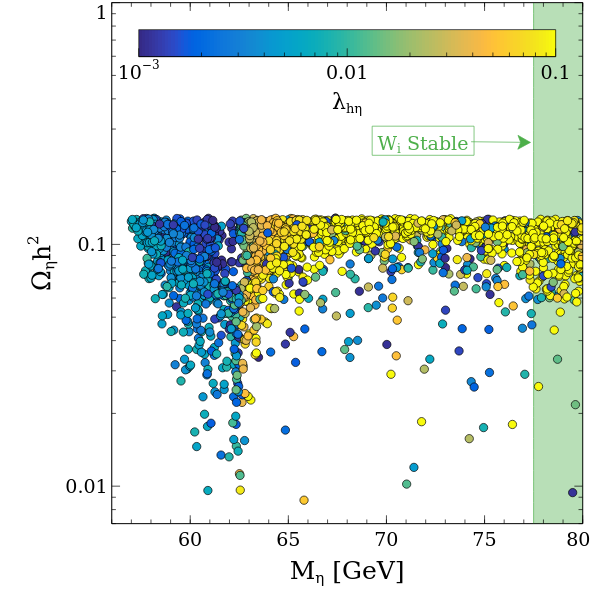}
\caption{LP shows the scatter plot in the $M_{\psi}-\Omega_{\psi + \chi}h^{2}$
planes whereas the RP shows the scatter plot in the 
$M_{\eta_{1}} - \Omega_{\eta_{1}}h^{2}$ plane. The colour bar in the 
LP shows the different values of $SU(2)_{D}$ gauge coupling and 
the colour bar in the RP shows the different values of the quartic 
coupling $\lambda_{h\eta}$.} 
\label{scatter-plot-1}
\end{figure}

In the LP of Fig. \ref{scatter-plot-1}, we draw the scatter
plot in the $M_{\psi}-\Omega_{\psi + \chi} h^{2}$ planes. The colour 
variation in the points shows the different values of $SU(2)_{D}$
gauge coupling $g_{D}$. We have kept the dark gauge boson mass
fixed at $M_{W_D} = 155$ GeV, therefore, we observe resonance
annihilation via $ W_D $ around $M_{\psi} = 77$ GeV. 
Near the resonance region, lower values of $g_{D}$ give higher values 
of the fermionic relic density while outside the resonance region, this density is mainly governed by the 
annihilation into gauge bosons channel and as expected, it varies inversely with the gauge coupling. 
We see that there is indeed a large part of the parameter space where the
fermionic relic density is too small and therefore an asymmetric component
could take over and give the correct DM abundance, as discussed in our
previous paper \cite{Biswas:2018sib}.

In the RP of Fig. \ref{scatter-plot-1}, we show the scatter plot in the $M_{\eta} - \Omega_{\eta} h^{2}$ 
plane where the colour variation displays the value of $\lambda_{h\eta}$. 
We do not observe any sharp resonance region near $M_{\eta} \sim \frac{M_{h_{1}}}{2}$ since the 
late decays 
$\psi^{\prime}, \chi^\prime \rightarrow \eta \nu$ 
also contribute substantially to the density. 
We also find that away from the resonance region, we need a larger value of the coupling $\lambda_{h\eta}$ to match the DM relic
density, as shown by the yellow points. Indeed, lower values of the couplings give a significantly large value of the DM density as shown 
in the RP of Fig. \ref{line-plot-4}, and are ruled out by the Planck upper bound on DM density.
\begin{figure}[h!]
\centering
\includegraphics[angle=0,height=7.5cm,width=8.5cm]{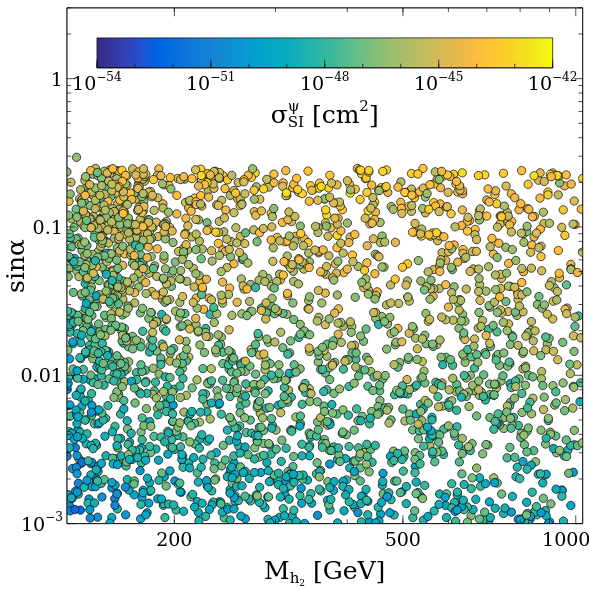}
\includegraphics[angle=0,height=7.5cm,width=8.5cm]{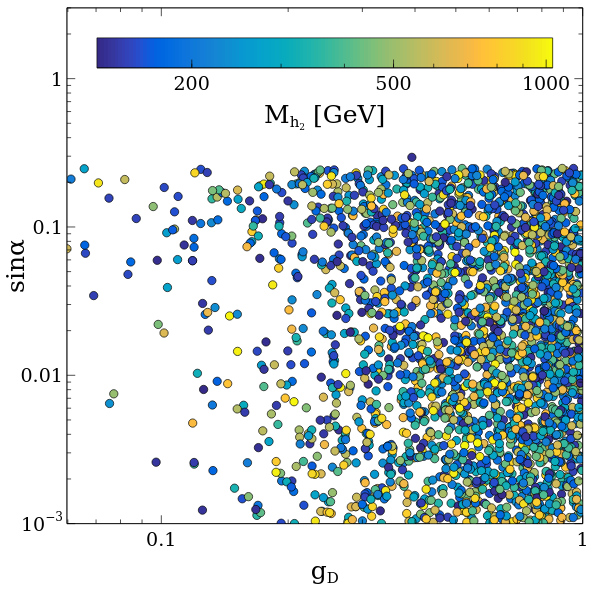}
\caption{LP and RP show the scatter plots in the $M_{h_{2}}-\sin\alpha$
and $g_{D}-\sin\alpha$ planes. The colour bar in the LP represents the
values of the SIDD cross-section whereas in the RP we have 
shown the second Higgs mass $M_{h_{2}}$. } 
\label{scatter-plot-2}
\end{figure}

In the LP and RP of Fig. \ref{scatter-plot-2}, we draw scatter
plots in the $M_{h_{2}}-\sin\alpha$ and $g_{D}-\sin\alpha$ planes. The
colour bar in the LP represents the SIDD cross-section for $\psi$ DM
whereas in the RP we have shown it for the BSM Higgs mass $M_{h_{2}}$.
In the LP, we find that $\sin\alpha > 0.23$ is not allowed due to constraints from the Higgs signal strength, along with additional 
bounds coming from BSM Higgs physics searches and the decay of SM Higgs into DM which appear as missing energy.

As mentioned earlier in Section \ref{model}, we have used \texttt{HiggsBound} and 
\texttt{HiggsSignal} packages to implement the bounds associated with
the Higgses. We also see a linear correlation between the mixing angle
$\sin\alpha$ and the SIDD cross-section $\sigma^{\psi}_{SI}$
as exhibited in Eq.~(\ref{SIDD-psi}).

 In the RP of Fig. \ref{scatter-plot-2}, we see that $g_{D} < 0.1$ is almost forbidden because it overproduces the 
 DM for most of the mass range except the resonance regions, as shown in Fig. \ref{line-plot-1}. 
 Therefore, we have very few points at low $ g_D $.
 From Fig. \ref{line-plot-4} one can see that $\sin\alpha$ affects the
 $ \psi$ relic density only at low mass, where the annihilation channels via gauge interactions are
 closed, and therefore we do not expect any correlation between $g_{D}$ and $\sin\alpha$, as seen in the figure.
Also, the value of $ M_{h_{2}} $ does not correlate with the couplings, as visible from the colours, because $M_{h_{2}}$ 
does not affect the DM relic density apart for shifting the $h_{2}$ resonance region as shown in Fig. \ref{line-plot-2}. 
\begin{figure}[h!]
\centering
\includegraphics[angle=0,height=7.5cm,width=8.5cm]{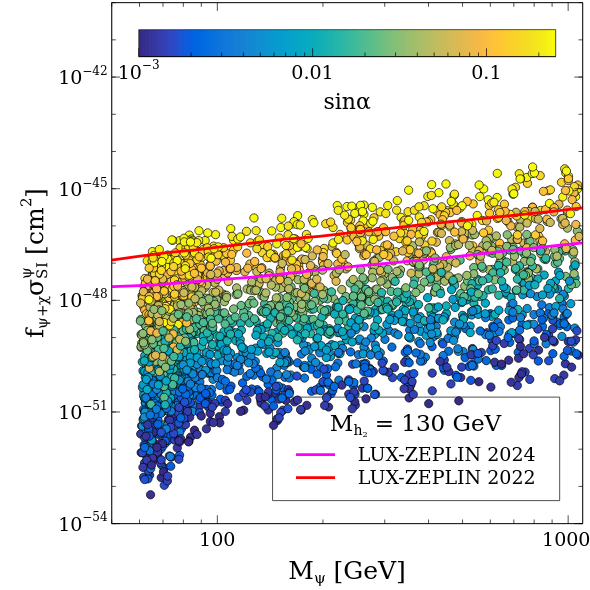}
\includegraphics[angle=0,height=7.5cm,width=8.5cm]{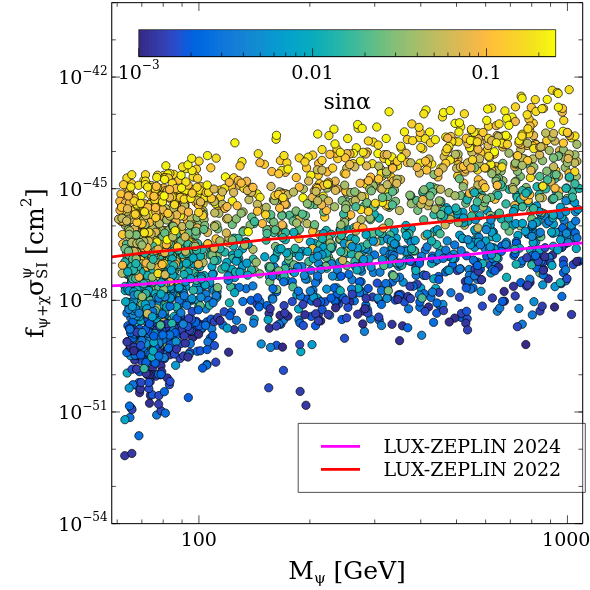}
\caption{LP and RP show the same plot $M_{\psi} - f_{\psi+\chi} \sigma_{SI}$
where for the we have fixed the $M_{h_{2}} = 130$ GeV and for RP
we have varied it. The colour bar in both plots represents the 
different values of the mixing angle $\sin\alpha$. In the RP, we have varied the $M_{h_2}$ mass
as shown in Eq. (\ref{range-of-parameter}).} 
\label{scatter-plot-3}
\end{figure}

In the LP and RP of Fig. \ref{scatter-plot-3}, we give the scatter plot in the 
$M_{\psi} - f_{\psi+\chi} \sigma^{\psi}_{SI}$ plane\footnote{$f_{A}$ is the fraction of the 
$A = \psi, \chi, \eta$ DM compare to the total
DM density {\it i.e.} 
$f_{A} = \frac{\Omega_{A} h^{2}}{0.12 }$.} 
where the strength of the mixing angle $\sin\alpha$ is shown in colour. 
We see the linear correlation between the effective SIDD cross section and 
the mixing angle, given explicitly by Eq. (\ref{SIDD-psi}). 
From the same Eq. (\ref{SIDD-psi}), we can also see that if $M_{h_{2}}$ and $M_{h_{1}}$ 
are similar then there is a suppression in the SIDD cross-section as evident in the LP of Fig. \ref{scatter-plot-3}, with $M_{h_{2}} = 130$ GeV. 
 It is to be noted that we can avoid the direct detection bounds by choosing $M_{h_{2}}$ close to 
 the SM Higgs mass $M_{h_{1}}=125$ GeV, at the cost of some fine-tuning.
 From the LP, we note that $\sin\alpha > 0.2$ is ruled
 out by the LUX-ZEPLIN data\cite{LZ:2022lsv, LZCollaboration:2024lux} 
 whereas in the RP, where we do not have any
 cancellation due to the mass of $M_{h_{2}}$, $\sin\alpha > 0.03$ is ruled 
 out by the recent LUX-ZEPLIN DD data.

\begin{figure}[h!]
\centering
\includegraphics[angle=0,height=7.5cm,width=8.5cm]{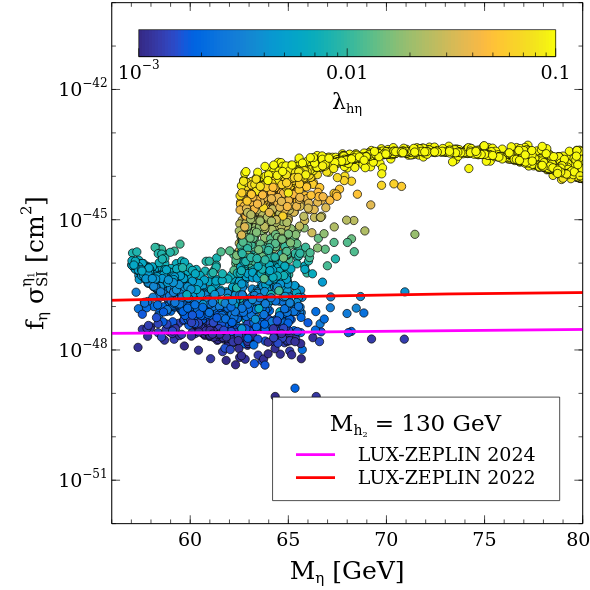}
\includegraphics[angle=0,height=7.5cm,width=8.5cm]{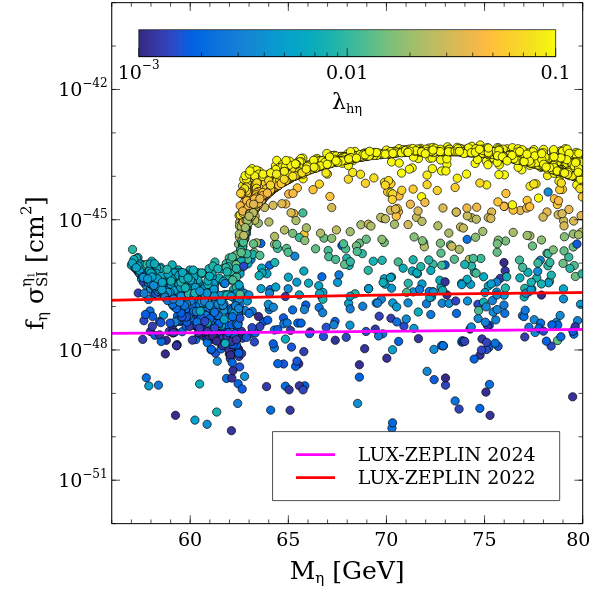}
\caption{Both the plots display the variation of the same plane 
$M_{\eta} - f_{\eta} \sigma_{SI}$. The LP is for the second Higgs
mass $M_{h_{2}} = 130$ GeV and the RP is for varied values of $M_{h_{2}}$.
The colour variation in both plots represents the different values
of $\lambda_{h\eta}$. Again in the RP, we have varied the $M_{h_2}$ mass
in the range shown in Eq. (\ref{range-of-parameter}).} 
\label{scatter-plot-4}
\end{figure}

The LP and RP of Fig. \ref{scatter-plot-4} give the scatter plots in the $M_{\eta} - f_{\eta} \sigma^{\eta_{1}}_{SI}$
planes where the colours show the different values of the portal coupling $\lambda_{h\eta}$. 
In the LP of Fig. \ref{scatter-plot-4}, we have shown the plot for $M_{h_{2}} = 130$ GeV whereas in the RP we have varied its mass. 
In the LP, when $M_{\eta} > 70$ GeV, there are no viable points for 
$\lambda_{h\eta} < 0.07$, while in the RP there exists no such bound on
$\lambda_{h\eta}$. The points that appear in the RP are due to
the second Higgs resonance in the subprocess $\eta \bar \eta
\xrightarrow[]{h_{2}} f \bar f$. Moreover, since the SIDD cross-section
for $\eta $ is directly proportional to $\lambda_{h\eta}^2$,
the RP has a few points below the current bound of LUX-ZEPLIN.
In the LP, due to the $h_2$ mass suppression in the DD cross-section, we obtain viable points below the LUX-ZEPLIN bound near the SM Higgs resonance, which overlaps with the second Higgs resonance. 
In contrast, in the RP, the mass of $h_2$ is varied which allows the resonance to move to higher values of the DM mass and helps in avoiding the bounds for any DM mass.
It is worth mentioning that we do not have a sharp resonance for
 $M_{\eta_{1}} \sim \frac{M_{h_{1}}}{2}$ because the decay modes
$\psi^{\prime}, \chi^{\prime} \rightarrow \eta \nu$ also
gives a substantial contribution to the $\eta$ DM density.
\begin{figure}[h!]
\centering
\includegraphics[angle=0,height=7.5cm,width=8.5cm]{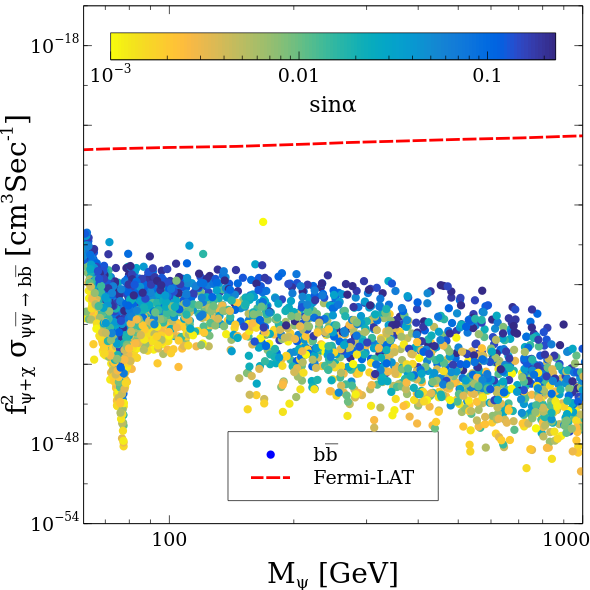}
\includegraphics[angle=0,height=7.5cm,width=8.5cm]{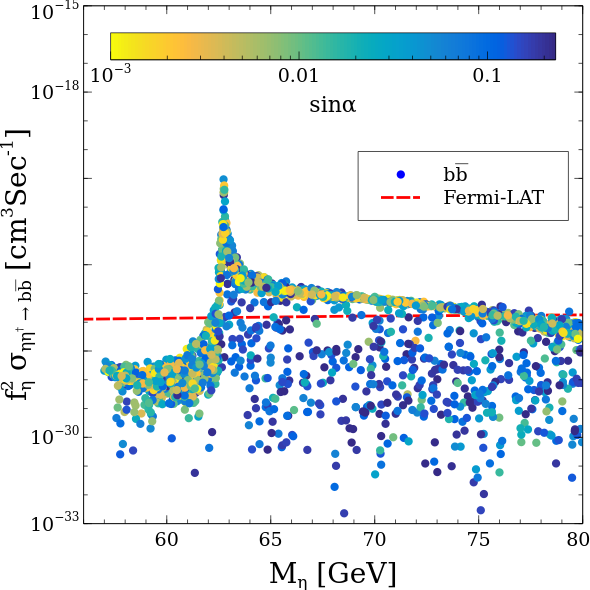}
\caption{LP (RP) shows the scatter plot in the $M_{\psi} 
- f^{2}_{\psi} <\sigma v >_{\psi \bar \psi \rightarrow b \bar b}$
($M_{\eta} 
- f^{2}_{\eta} <\sigma v >_{\eta \bar\eta \rightarrow b \bar b}$) plane. 
The colour bar displays the different values of the mixing 
angle $\sin\alpha$. The red dashed line represents the bound from the 
Fermi-LAT data \cite{MAGIC:2016xys}.} 
\label{scatter-plot-5}
\end{figure}

In Fig. \ref{scatter-plot-5}, we study the indirect detection prospects for
the DM candidates $\psi$ and $\eta$. To compute the indirect detection
cross-sections, we use the expressions shown in 
Eqs. (\ref{etaetabb}, \ref{psipsibb}) and the fractions
$f_{\psi+\chi, \eta}$ are obtained using \texttt{micrOmega}.  
In the LP of Fig. \ref{scatter-plot-5}, we observe that the indirect detection cross-section 
is below the current limit from Fermi-LAT, because for $\psi$ DM, 
the dominant annihilation 
modes are the invisible $\psi \bar \psi \rightarrow W_D W_D $, $\psi \bar \psi \rightarrow \eta \eta$ as well as $\psi \bar \psi \rightarrow 
b \bar b$ cross section times velocity is suppressed by the square 
of the dark matter velocity.
Moreover, we see a linear relation between the indirect detection cross-section and the mixing angle 
$\sin\alpha$ because $\psi$ DM interaction with the visible sector is always suppressed by the mixing angle $\sin\alpha$.
In the RP, we show the indirect detection prospects for the 
$\eta$ DM candidate. Here we do not see a suppression in the
density near the SM Higgs resonance due to the DM production coming from 
the late decay of $\psi^{\prime}, \chi^{\prime}$ to the $\eta$. 
A non-negligible portion of the parameter space at the SM Higgs 
resonance is already ruled out by indirect detection. Note that  $\eta$ 
can directly interact with the visible sector by quartic coupling $\lambda_{h\eta}$,
so the indirect detection signal does not depend only on the 
mixing angle $\sin\alpha$.
Therefore, we do not expect any correlation between indirect detection rates
and the mixing angle as also seen from the figure.

We conclude from the previously discussed figures that the direct detection signal can be obtained from 
both DM candidates and could be within the reach of the next generation of DD experiments, apart 
from a tuned cancellation. On the other hand, only the scalar DM component can give an observable ID signal in the near future.

\section{Collider Analysis}

\begin{figure}[h!]
\centering
\includegraphics[angle=0,height=7.5cm,width=8.5cm]{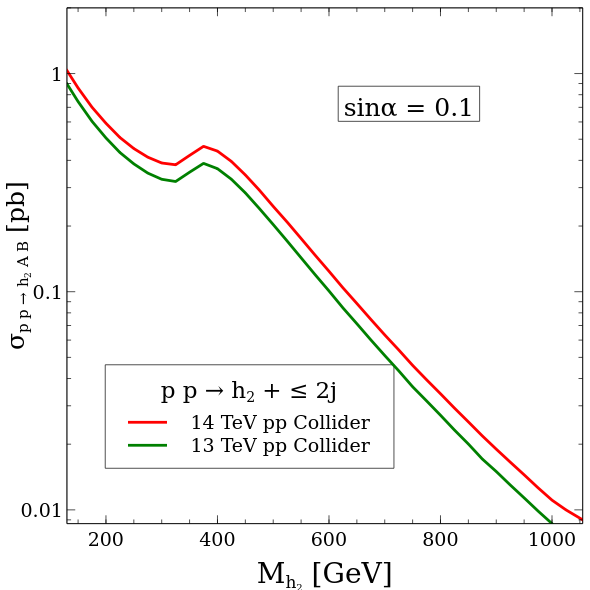}
\includegraphics[angle=0,height=7.5cm,width=8.5cm]{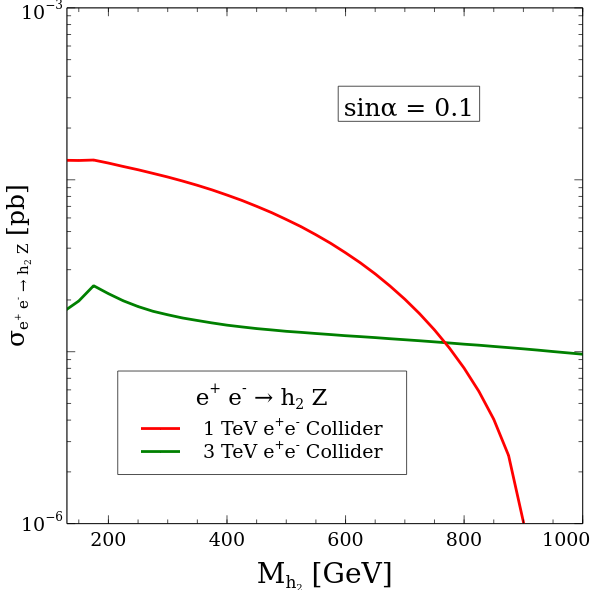}
\caption{LP: Production cross-section of $pp \rightarrow h_{2} + \leq 2j$ 
at the 14 TeV and 13 TeV $pp$ collider. In the RP, we have 
$e^{+}e^{-} \rightarrow h_{2} Z$ production cross section for 1 TeV and 
3 TeV run of the $e^{+}e^{-}$ linear collider. In both the plots,
we have kept the scalar mixing angle fixed at $\sin\alpha = 0.1$. In the LP, we
have considered $p^j_{T} > 20$ GeV and $\eta_{j} < 5$.} 
\label{production-CS}
\end{figure}

In this section, we discuss the collider prospects of production
and detection for the BSM Higgs $h_{2}$. There have been several studies of BSM Higgs or scalar that can be produced through its mixing with the SM sector \cite{ATLAS:2017uhp, ATLAS:2017tlw, ATLAS:2017ayi, Belanger:2021slj, Banerjee:2015hoa, Banerjee:2016foh}.
In our model, there are very different decay channels for the BSM Higgs 
ranging from the dominant decay mode to the SM sector 
to mainly decay
to the dark sector and therefore the phenomenology is very rich.
In this work, we explore the region of the parameter space where the BSM 
Higgs $h_{2}$ mainly decays to the dark sector including the DM
candidates.  We focus on the search for events with multi-jet plus missing
energy ($\geq 2j + \cancel{E}_{T}$) using the cut-based analysis.  

In Fig. \ref{production-CS}, we show the production cross-section of $h_{2}$ at the $pp$ and $e^{+}e^{-}$ colliders 
in association with jets and gauge boson $Z$, respectively. As we highlight in the next figure, $h_{2}$ mainly 
decays to stable DM particles if the processes are kinematically allowed, leading to missing energy signals associated with additional jets. 
The production of $h_{2}$ in association with the jets in the 
final states opens up the possibility of studying either 
mono-jet plus missing energy ($1j+\cancel{E}_{T}$) or 
multi-jet plus missing energy signal ($\geq2j+\cancel{E}_{T}$). 
We find that the $\geq2j+\cancel{E}_{T}$ signal is more promising compared to the $1j+\cancel{E}_{T}$ signal. We therefore present
our results for $\geq 2 j + \cancel{E}_{T}$ signal by considering
all the relevant SM backgrounds using the cut-based analysis. 

On the other hand, the production cross-section of $h_{2}$ at an electron
collider is too small and we do not expect a sufficient rate for detection.
Note that the mixing angle $\alpha$ that determines the production rate is
already bounded by Higgs data to be $\sin\alpha \leq 0.23$ as shown
in the LP of Fig. \ref{scatter-plot-2}. Therefore, we refrain from any further 
discussion of its detection prospects at the $e^{+}e^{-}$ collider.
\begin{figure}[h!]
\centering
\includegraphics[angle=0,height=7.5cm,width=8.5cm]{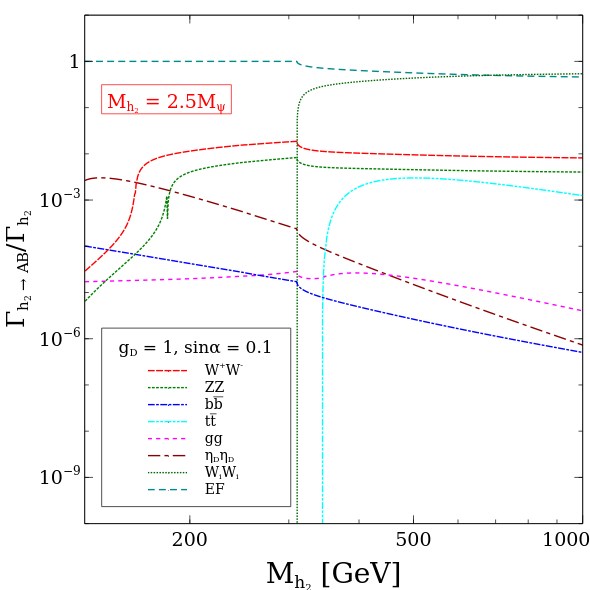}
\includegraphics[angle=0,height=7.5cm,width=8.5cm]{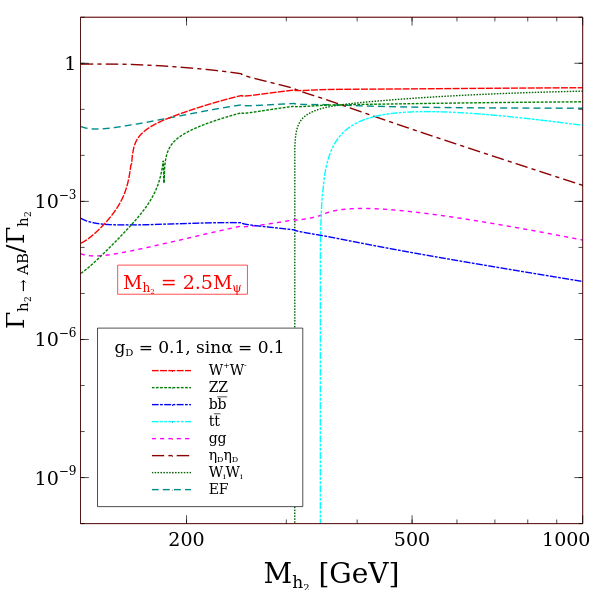}
\includegraphics[angle=0,height=7.5cm,width=8.5cm]{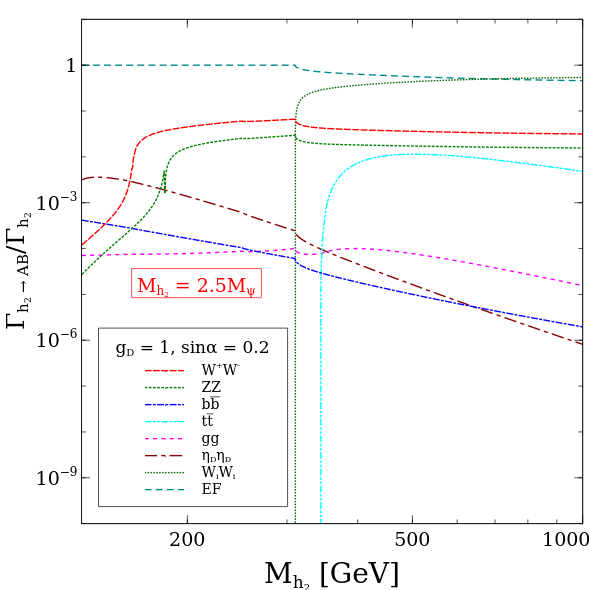}
\includegraphics[angle=0,height=7.5cm,width=8.5cm]{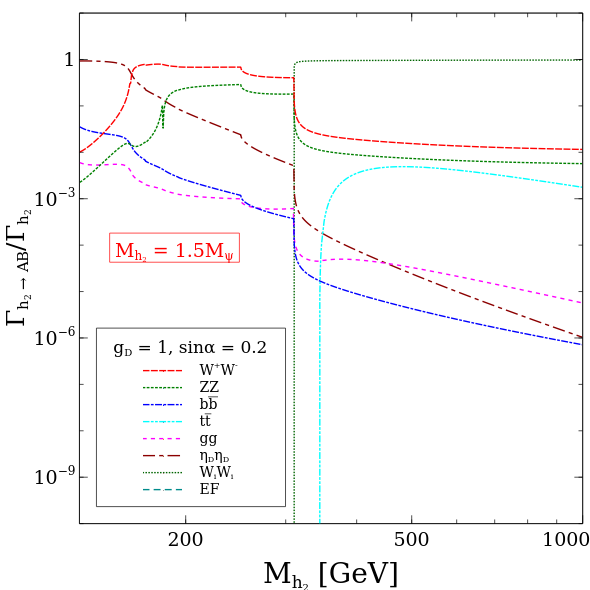}
\caption{Variation of beyond SM Higgs decays to different channels for
different values of $h_{2}$ mass. In generating the plots, other parameters 
have been kept fixed at $M_{\eta_1} = M_{\eta_2} = 62$ GeV,
$M_{W_i} = 155$ GeV ($i = 1, 2, 3$), 
$\lambda_{hD} = \lambda_{h\eta} = \lambda_{D1} = 10^{-2}$. In the plots, EF is the exotic fermions
corresponds to $\psi, \chi, \psi^{\prime}, \chi^{\prime}$.} 
\label{H2-branching}
\end{figure}

In Fig. \ref{H2-branching}, we plot the branching fractions for the decay
of the BSM Higgs $h_{2}$ to different final states, including SM and BSM particles. Since the model contains many states lighter than $h_2$, many channels are possible depending on the mass spectrum. 
For example, when $M_{h_{2}} < 2 M_{W_D}$, the dominant decay mode of $h_{2}$ is to  
$\psi \bar{\psi} + \chi \bar{\chi} + \psi^{\prime} \bar{\psi^{\prime}} 
+ \chi^{\prime} \bar{\chi^{\prime}}$ if it is kinematically 
allowed.
%\footnote{The branching for the $\psi^{\prime}$, $\chi^{\prime}$ will be similar to that of $\psi$, %$\chi$ as they have the same mass and have not been displayed in the plots.}. 
Once the $h_{2} \rightarrow W_D W_D $ opens up, it becomes the dominant decay mode similar to what happens in the case of the SM
Higgs decaying to the EW gauge bosons.
In the plots, we have considered different values of $g_{D} $ and $\sin\alpha $, as well as different ratios of the $ h_2 $ mass to the DM fermion mass $ M_\psi$,
while the other parameters have been kept fixed as given in the caption of the figure. 
As can be seen from the panels, the BSM Higgs mostly decays to stable invisible 
particles and it will produce a missing energy signal at colliders. 
The top-right plot has been generated for $g_{D} = 0.1$ and the other parameters have been kept fixed as in the top-left plot. 
Due to the lower value of $g_{D}$, the branching ratios
of $h_{2} \rightarrow \psi\bar{\psi}, \chi\bar{\chi}, W_D W_D $
decrease, as can be easily understood from the analytical expressions
given in Eqs. (\ref{H2-psi-psi}), (\ref{H2-WiWi}). On the contrary, the
$h_{2} \rightarrow \eta \bar \eta $ branching ratio increases since it is 
proportional to the v.e.v. $v_{D}$, and inversely proportional
to $g_{D}$ at fixed $W_D$ mass. 
This result is in full agreement with Eq. (\ref{H2-ss}).
In the bottom-left plot, we have increased the mixing angle to $\sin\alpha = 0.2$ and 
this simply increases the production rate but does not seem to modify the branching ratios too much,
apart from a slight increment in the $W^{+}W^{-}, ZZ$ channels. 
In the bottom-right plot, we have instead chosen $M_{h_{2}} < 2 M_{\psi,\chi}$, while
keeping the rest of the parameters as before. Then the decay channel into the DM fermions is closed and the SM channels are dominant until the decay into $W_D W_D$ opens up. So, with this particular choice of $h_{2}$ mass, we obtain a parameter space where $h_2$ has dominant SM decay modes and could look like a second SM Higgs. 

The analytical expressions for the $h_2$ decays into BSM particles are presented in Appendix \ref{H2-decay-expression}, 
while the $h_2$ decay modes into SM are obtained by rescaling the SM Higgs decay rates with the mixing angle. 
We observe that due to the small mixing angle between 
the SM and dark Higgs, the dominant decay, if kinematically open, is to stable (or metastable) dark particles, which manifest 
as missing energy at colliders. Note that the branching ration of $h_{2}$ to 
$\eta \bar \eta $ is often very small, even when these states are light since we have chosen small quartic couplings. 
As long as the fermionic DM channel dominates, 
the phenomenology of the model at colliders, with symmetric DM, is similar to the scenario with asymmetric DM studied in Ref. \cite{Biswas:2018sib}. 
In fact the production and decay of $h_2$ can proceed in the same way in that scenario,
where the decay of $h_2$ into $\eta \bar \eta $ is kinematically closed.
The BSM Higgs signal in this model turns out to be different from previous BSM Higgs searches given in the literature, which mostly focus on the $h_{2}$ decays to the SM gauge bosons $W^{+}W^{-}$ and $ZZ$ \cite{ATLAS:2020tlo}. We can realise this scenario also in our model, but only if the fermions are heavier than half the BSM Higgs. 

{\bf Simulation details:}
To study the detection prospect for the  BSM Higgs ($h_{2}$) at the 14 TeV run of the LHC, we have generated our events using the MadGraph event generator for both the signal and the main SM backgrounds.
As pointed out in Fig.\,(\ref{H2-branching}), our BSM Higgs mostly decays to the dark sector particles which manifest at the detector only as the missing energy. Therefore, to identify the large missing energy signal we have to rely on producing some visible states that recoil against the missing particles which in this case are the jets. This will allow us to look for multi-jet plus missing energy signal at the experiment. To determine the rate and the shape of the signal, we have simulated the production and decay of the BSM Higgs at the 14 TeV run of the LHC through the following subprocesses,
\begin{eqnarray}
p\, p &\rightarrow& h_{2} \nonumber \\
p\, p &\rightarrow& h_{2}\,j \nonumber \\
p\, p &\rightarrow& h_{2}\,j\,j \nonumber 
\end{eqnarray}   
The events have been generated using the {\tt Madgraph} (V3.4.1) 
\cite{Alwall:2014hca, Alwall:2011uj} event generator with the
{\tt CTEQ6L} \cite{Pumplin:2002vw} as the parton distribution function. 
The generated events are then passed through {\tt Pythia8} \cite{Bierlich:2022pfr} for showering and hadronization. The 
partons and showered jets are matched using the MLM matching scheme 
\cite{Mangano:2006rw, Hoeche:2005vzu}. The generated event files from Pythia (STDHEPMC extension) are then fed into {\tt Delphes} \cite{deFavereau:2013fsa, Selvaggi:2014mya, Mertens:2015kba} for object reconstruction and detector simulation, and for jet formation using anti-$k_T$ algorithm \cite{Cacciari:2008gp}
with {\tt Fastjet} \cite{Cacciari:2011ma}. For detector simulation, we have used the default CMS card available in {\tt Delphes} and modified it to include the stable particle content as needed for our model.

Here we aim to study missing energy associated with a multi-jet signal in 
the final state {\it i.e.} more explicitly  the final state contains 
$\geq 2j + \cancel{E}_{T}$. 
SM backgrounds that can produce identical final states include 
pure QCD jets production, $W$+jets, $Z$+jets, $t\bar{t}$+jets, $WZ$+jets and $ZZ$+jets.
As we shall see later, the signal rate is tiny compared to the huge SM backgrounds. Therefore, we need to devise suitable cuts for the different collider kinematical variables to reduce the SM backgrounds significantly without loosing too much of the signal. The basic cuts that we have imposed to reduce QCD and EW backgrounds are 
\begin{itemize}
\item[A0:] 
\begin{itemize}
\item[-] All charged leptons must have a minimum transverse momentum 
$P^{l_i}_{T} > 20$ GeV with an upper bound on their pseudorapidity as
$|\eta_{l_i}| < 2.5$.
\item[-] Photons are identified when their transverse momentum 
$p^{\gamma}_{T} > 20$ GeV and the pseudorapidity $\eta^{\gamma} < 2.5$.
\item[-] The jets are selected if their transverse momentum 
$p^{j_i}_{T} > 20$ GeV and remain central with the maximum rapidity 2.5 {\it i.e.} $\eta_{j_i} < 2.5$.
\item[-] The azimuthal separation between jets and missing energy vector
must be greater than 0.2 {\it i.e.} 
$\Delta \phi(j_i,\cancel{E}_{T}) > 0.2$. 
\end{itemize}
\end{itemize}      
To reduce the backgrounds further, we exploit the following cuts that do not suppress the signal. The most commonly used kinematical variables in minimising the backgrounds include,
\begin{itemize}
\item[$\rightarrow$] first is the scalar sum of the transverse momentum of
the jets, defined as,
\begin{eqnarray}
H_{T} = \sum_{i \in jets} P^{j_i}_{T}\,.
\label{HT-variable}
\end{eqnarray}
\item[$\rightarrow$] second in the list is the effective mass which is 
the sum of the transverse momentum of the jets, leptons and missing energy
and defined as,
\begin{eqnarray}
M_{Eff} = \sum_{i \in jets} P^{j_i}_{T} 
+ \sum_{i \in leptons} P^{l_i}_{T}
+ \cancel{E}_{T}
\end{eqnarray} 
\item[$\rightarrow$] the other relevant kinematical variables which are also
important in our analysis to minimize the backgrounds are the ratio
of transverse missing energy and previously defined kinematical variables
which are defined as,
$R_{1} = \frac{\cancel{E}_{T}}{M_{Eff}}$ and $R_{2} = \frac{\cancel{E}_{T}}{H_{T}}$\,.
\end{itemize}   
\begin{figure}[h!]
\centering
\includegraphics[angle=0,height=7.5cm,width=8.5cm]{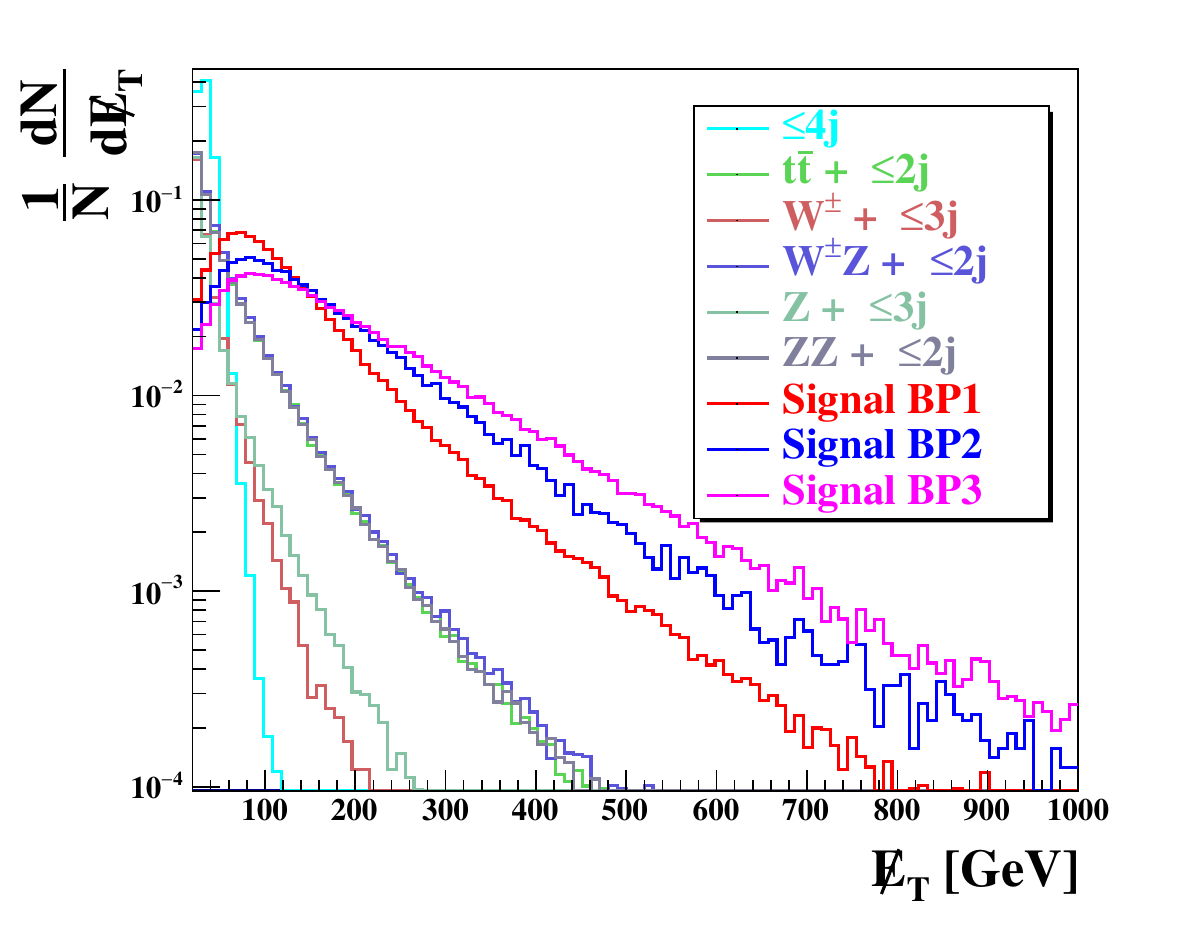}
\includegraphics[angle=0,height=7.5cm,width=8.5cm]{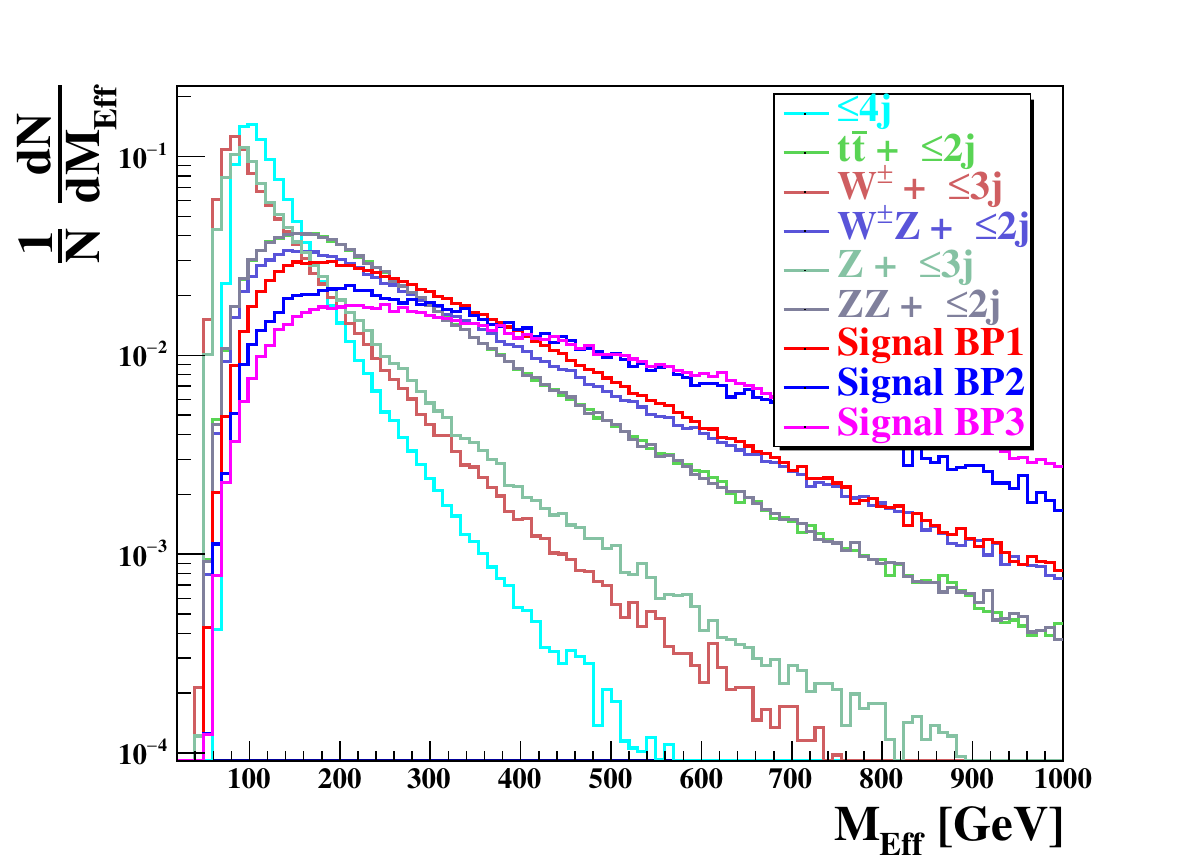}
\caption{Histograms in the LP and RP which show the variation of the fraction of the total  
number of the events for the signal and SM backgrounds w.r.t 
 kinematical variables $\cancel{E}_{T}$ and $M_{eff}$, respectively, in a
 fixed bin size.} 
\label{hist-1}
\end{figure}

\begin{figure}[h!]
\centering
\includegraphics[angle=0,height=7.5cm,width=8.5cm]{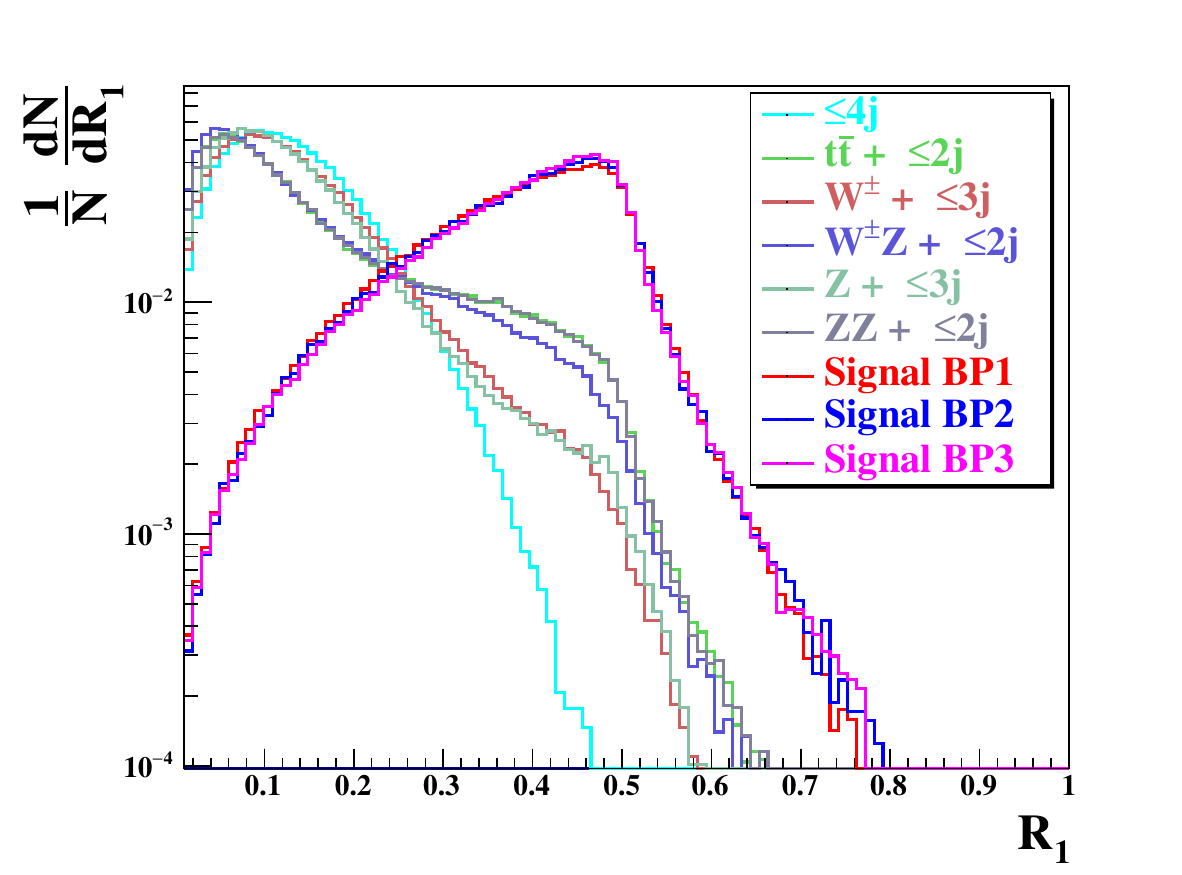}
\includegraphics[angle=0,height=7.5cm,width=8.5cm]{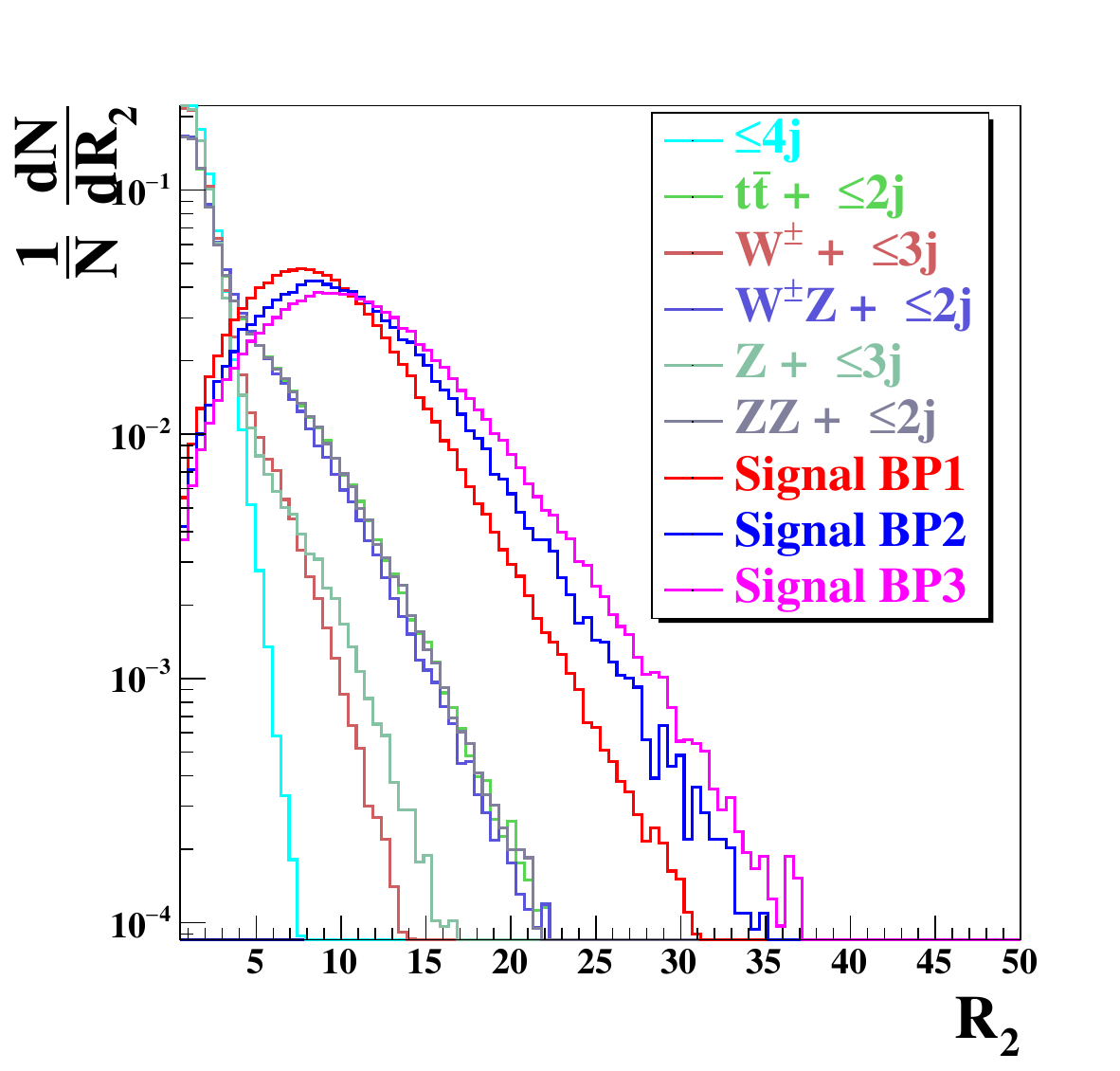}
\caption{Histograms in the LP and RP which show the variation of the fraction of the total number of events in the fixed bin size of 
the kinematical variables  
$R_{1}$ and $R_{2}$, respectively.} 
\label{hist-2}
\end{figure}

We see in the histograms plots in Fig.\,\ref{hist-1} and \ref{hist-2} 
that the signal dominates the events at large missing energy and
$ M_{Eff} $, as well as for large $ R_1,R_2$ values which allows 
to discriminate between signal and background. 
With a suitable choice of cuts on the kinematical variables, we can 
suppress the background significantly. 

Therefore, we impose the following cuts on the variables discussed above to optimize the signal-to-background ratio.
\begin{itemize}
\item[A1:] We select our events for the final states which contains 
$\geq 2j + \cancel{E}_{T}$. Since we are studying hadronic states 
 in the final state, we vetoed the charged leptons, photons, and b-jets.
 \item[A2:] We have considered relatively hard cuts on the transverse
 momentum of the leading and next-to-leading jets which are 
 $p^{j_1}_{T} > 130$ GeV and $p^{j_2}_{T} > 80$ GeV, respectively.
 \item[A3:] The azimuthal separation between the jets and missing energy
 for the leading and next to leading jet must be greater than 0.4
 {\it i.e.} $\Delta \phi(j_{1,2},\cancel{E}_{T}) > 0.4$ and rest of the    
jets satisfy $\Delta \phi(j_{i},\cancel{E}_{T}) > 0.2$.
\item[A4] We have demanded that $R_{1} > 0.3$ and $R_{2} > 15$. These cuts
are very effective in reducing the backgrounds, as can be seen in the histograms.
\item[A5:] The bound on the effective mass has been taken as $M_{Eff} > 800$ GeV in our analysis.
\item[A6:] Finally, we require the missing transverse energy $\cancel{E}_{T} > 300$ GeV.
\end{itemize}  

As we observe later, the A5 cut turns out to be very efficient in reducing all SM background contributions, 
and because of the very hard cut on $M_{Eff}$ which includes the $\cancel{E}_{T}$ in its definition, the cut 
on the missing transverse energy (A6) is not very effective. 
Moreover, the cut A4 is also dependent on the value of $H_{T}$ and $M_{Eff}$, which are correlated with the cut A5. 

We will now discuss the signal and background survival after imposing
the above mentioned cuts.
In Table \ref{sig-tab1}, we present three benchmark points that can 
account for the total DM density, with contributions from both fermion 
and scalar DM candidates, as discussed earlier.
%\color{black}
%We focus on the detection of our signal at the 14 TeV run of LHC, while the relevant SM background processes are $W^{\pm} + \leq 3j$, $Z + \leq 3j$, $\leq 4j$, $WZ+\leq 2j$, $t\bar{t} + \leq 2j$, $ZZ+\leq 2j$. 
We compute the effective cross-sections for the signal and background
at the 14 TeV run of LHC and the effect of the various cuts on their rates, and list them in Tables \ref{sig-tab1} and \ref{bkg-tab2}.

\def\I{i}
\begin{center}
\begin{table}[h!]
%\begin{center}
\begin{tabular}{||c|c||}
%\topline
\hline
\hline
\begin{tabular}{c|c|c}
    %\hline
    \multicolumn{3}{c}{Signal at 14 TeV}\\ 
    \hline
    BP & $M_{h_2}$ [GeV]&Cross-section (fb) \\ 
    \hline
    BP1 &200 &459.29\\ 
    \hline
    BP2 &400& 181.76 \\
    \hline
    BP3 &600& 37.60\\
\end{tabular}
&
\begin{tabular}{c|c|c|c|c|c}
    %\hline    
    \multicolumn{6}{c}{Effective Cross section after different cuts (fb)}\\   
    \hline 
    A0\,+\,A1 & \,\,~A2~\,\, &\,\, A3\,\, &\,\, A4\,\, &\,\, A5 \,\, &\,\, A6 \\
    \hline
    201.14 & 29.96 & 28.09 & 7.39 & 4.16 & 4.16  \\
    \hline
    102.07 & 25.23 & 23.57 & 7.74 & 4.84 & 4.84  \\
    \hline
    23.65 & 7.32 & 6.83 & 2.65 & 1.87 & 1.87   \\
\end{tabular}\\
\hline
\hline
\end{tabular}
%\bottomrule
%\hline
%\hline
\caption{Cut-flow table of the signal production cross section at 14 TeV
$pp$ collider for the three benchmark points after applying all 
the cuts. The other relevant parameters are kept fixed at 
$M_{\eta}  = 62$ GeV,
$M_{\psi, \chi} = 66.5$ GeV, $M_{\psi^{\prime}, \chi^{\prime}} = 500$ GeV,
$M_{W_D} = 155$ GeV, $g_{D} = 1$, $\sin\alpha = 0.1$, 
$\lambda_{hD} = \lambda_{h\eta} = \lambda_{D1} = 10^{-2}$. 
We consider a K-factor equal to 3.}
\label{sig-tab1}
%\end{center}
\end{table}
\end{center}
%%%%%%%%%%%%%%%%%%%%%%%%%%%%%%%%%%%

%%%%%%%%%%%%%%%%%%%%%%%%%%%%%%%%%%%
%%%%%%%%%%%%%%%%%%%%%%%%%%%%%%%%%%%
\def\I{i}
\begin{center}
\begin{table}[h!]
%\begin{center}
\begin{tabular}{||c|c||}
%\topline
\hline
\hline
\begin{tabular}{c|c}
    %\hline
    \multicolumn{2}{c}{SM Backgrounds at 14 TeV}\\ 
    \hline
    Channels & Cross-section (pb) \\ 
    \hline
    Z $+\leq$ 3 jets & 5.59$\times 10^4$ \\ 
    \hline
    W$^{\pm}$ + $\leq$ 3 jets & 1.07$\times 10^5$  \\
    \hline
    QCD ($\leq$ 3 jets) & 1.91$\times 10^{8}$\\
    \hline
    t\,$\bar{\rm t} + $ $\leq$ 2 jets & $1.40\times10^{3}$\\
    \hline
    W$^{\pm}$Z + $\leq$ 2 jets & 55.35 \\
    \hline
    Z\,Z +$\leq$ 2 jets & 16.06\\ 
    \hline
    Total Backgrounds & ~\\ 
\end{tabular}
&
\begin{tabular}{c|c|c|c|c|c}
    %\hline    
    \multicolumn{6}{c}{Effective Cross section after different cuts (pb)}\\   
    \hline 
    A0\,+\,A1 & \,\,~A2~\,\, &\,\, A3\,\, &\,\, A4\,\, &\,\, A5\,\, &\,\, A6   \\
    \hline
    1.73 $\times 10^{3}$ & 697.37 & 459.68 & 5.15 & 2.68 & 2.68  \\
    \hline
    3.18 $\times 10^{4}$ & 1.27$\times10^{3}$ & 788.03 & 2.39 & 1.39 & 1.39 \\
    \hline
    8.17 $\times 10^{7}$ & 9.56 $\times 10^{5}$ & 6.07 $\times 10^{5}$ & - & - & -  \\
    
    \hline
    667.44 & 96.19 & 65.16 & 2.32 & 1.14 & 1.14 \\
    \hline
    28.10 & 6.35 & 4.28 & 0.09 & 0.05 & 0.05 \\
    \hline
    7.63 & 1.10 & 0.75 & 0.02 & 0.01 & 0.01  \\
    \hline
    ~ & ~ & ~ & ~ & ~ & 5.27 \\
\end{tabular}\\
\hline
\hline
\end{tabular}
%\bottomrule
%\hline
%\hline
\caption{Cut-flow table of SM backgrounds at the 14 TeV $pp$ collider
after applying all the cuts. The relevant cuts A0-A6 are elaborately described in the text.}
\label{bkg-tab2}
%\end{center}
\end{table}
\end{center}

 We see that the cuts are very effective in reducing the background 
 without losing too many signal events. In particular, the cuts A4 and 
 A5 brings down very effectively the dominant backgrounds like 
 $Z+\leq3j$, $W^{\pm}+\leq3j$ and $t \bar{t} +\leq 2j$.

After having simulated the signal and total background with all kinematic 
cuts, we can determine the statistical significance of the signal as
\begin{eqnarray}
\mathcal{S} = \sqrt{2 \times \left[ (s + b)\,{\rm ln}\left( 1 + \frac{s}{b}\right) -s \right]}.
\label{eq-significance}
\end{eqnarray}

In Table \ref{stat-tab3}, we give the 
statistical significance of the multi-jet + missing transverse energy signal for the three benchmark points for two values of the mixing angle, $\sin\alpha = 0.1$ and $0.2$. 
As the mixing angle enters quadratically in the production cross-section
for $ h_{2} $, a large mixing angle increases the signal.
For $\sin\alpha = 0.2$ and $M_{h_2} = 200$ GeV, the BSM Higgs may be observed at the 14 TeV run of LHC with 171 $fb^{-1}$ 
integrated luminosity at the $3\sigma $ level. The significance could reach as high as 12 $ \sigma $ for the high luminosity phase of the LHC with $3$ ab$^{-1}$. 
For higher masses or smaller mixing angles, more luminosity is needed to detect the BSM Higgs. However, we note that at the future run of LHC, the signal of multi-jet +  $\cancel{E}_{T}$ can achieve significances of about $3\sigma $.

%%%%%%%%%%%%%%%%%%%%%%%%%%%%%%%%%%%%%%%%%%%%%%%%%%%%%%%%%%%%%%%%%%%%%%%%%%%%%%%%%%%%%%%
\def\I{i}
\begin{center}
\begin{table}[h!]
%\begin{center}
\begin{tabular}{||c|c|c||}
%\topline
\hline
\hline
\begin{tabular}{c}
Signal\\
%\hline

\begin{tabular}{c|c}
    %\hline
    \multicolumn{2}{c}{ at 14 TeV }\\ 
    \hline
    BP & $h_{2}$ mass [GeV]\\ 
    \hline  
    BP1 & 200.0\\ 
    \hline
    BP2 & 400.0\\
    \hline
    BP3 & 600.0\\
\end{tabular}
\end{tabular}
& \begin{tabular}{c}
Statistical Significance ($\mathcal{S}$)\\
\hline
\begin{tabular}{c|c}
    %\hline
    \multicolumn{2}{c}{ $\mathcal{L} = 3000$ ${\rm fb^{-1}}$ }\\ 
    \hline
    $\sin\alpha = 0.1$ & $\sin\alpha = 0.2$\\ 
    \hline  
    3.14 & 12.55\\ 
    \hline
    3.65 & 14.40\\
    \hline
    1.41 & 5.64\\
\end{tabular}
\end{tabular}
&
\begin{tabular}{c}
Required Luminosity $\mathcal{L}$ (${\rm fb^{-1}}$)\\
\hline
\begin{tabular}{c|c}
    %\hline
    \multicolumn{2}{c}{ $\mathcal{S} = 3\sigma$ }\\ 
    \hline
    $\sin\alpha = 0.1$ & $\sin\alpha = 0.2$\\ 
    \hline  
    2741.45 & 171.48\\ 
    \hline
    2025.33 & 126.70\\
    \hline
    13565.0 & 848.12\\
\end{tabular}
\end{tabular}\\
\hline
\hline
\end{tabular}
%\bottomrule
%\hline
%\hline
\caption{Statistical significance of three benchmark points for 
$3000\,\,fb^{-1}$ luminosity data at the 14 TeV run LHC and 
the luminosity required for 
obtaining $3\sigma$ statistical significance of signal over background.}
\label{stat-tab3}
%\end{center}
\end{table}
\end{center}

\begin{figure}[h!]
\centering
\includegraphics[angle=0,height=7.5cm,width=8.5cm]{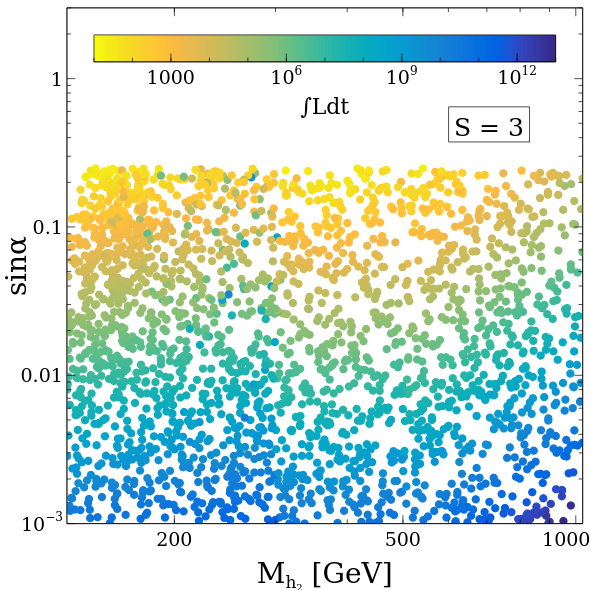}
\includegraphics[angle=0,height=7.5cm,width=8.5cm]{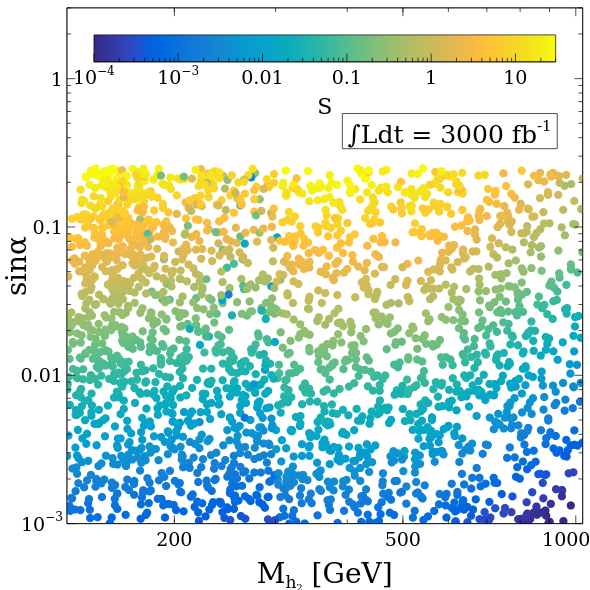}
\caption{LP: Scatter plot in the $M_{h_{2}}-\sin\alpha$ plane where the colour shows
the luminosity needed to obtain the statistical significance $\mathcal{S} = 3$. 
RP: The scatter plot again in the same plane but the colour variation represents
the different values of statistical significance $\mathcal{S}$ for integrated
luminosity $\int L dt = 3000\,\,fb^{-1}$. } 
\label{mh2-alpha-S-L}
\end{figure}

Now, let us go beyond the benchmark points and more generally explore the parameter space of the model. Since the mass of the BSM Higgs and the mixing angle determine the production cross-section, we have illustrated the prospects for detection in the $M_{h_{2}} - \sin \alpha$ in Fig. \ref{mh2-alpha-S-L}.
%%%%
%%%
 In the LP, the colour indicates the integrated luminosity $\int L dt$ required to obtain the statistical significance 
 $\mathcal{S} = 3$ for the signal $\cancel{E}_{T} + \geq 2j$, whereas in the RP, the color gives the statistical 
 significance $\mathcal{S}$ for a fixed value of integrated luminosity $\int L dt = 3000$ fb$^{-1}$. 
 In both plots, in determining the luminosity for a fixed value of statistical significance and vice versa, 
 we have considered an interpolated efficiency factor to determine the effective signal after applying all the cuts, 
 whereas the SM background is unaltered. We have used an interpolated value because the signal efficiency factor, 
 $\kappa = \frac{\rm signal\,\, CS\,\, after\,\, cuts}{\rm signal\,\, CS\,\, before\,\, cuts}$, is not the same for all the BPs 
 as shown in Table \ref{sig-tab1}. In both plots, all points satisfy the DM relic density bound as shown in 
 Eq. (\ref{DM-relic-density}), direct detection bound, indirect detection bound, and the bounds from collider experiments 
 on Higgses using the HiggsBound and HiggsSignal packages. 
 
 In the LP Fig. \ref{mh2-alpha-S-L}, we observe the variation in the $M_{h_2} - \sin\alpha$ plane, where $\sin\alpha > 0.23$ is disallowed by the SM Higgs signal strength. For a fixed value of $M_{h_{2}}$, as we increase $\sin\alpha$, the production cross-section rises, requiring lower luminosity to achieve $3\sigma$ statistical significance. Furthermore, as we explore larger values of $M_{h_{2}}$, higher luminosity is required because the production cross-section decreases with increasing $h_{2}$ mass, $M_{h_{2}}$. The figure indicates that the 14 TeV run of the LHC can access only a very limited range of $\sin\alpha$ and mass $M_{h_{2}}$, represented by the yellow points for 3000 $fb^{-1}$ luminosity data. In the RP, where significance is denoted by colour, it is evident that for $M_{h_2} \sim 200$ GeV and $
\sin\alpha \sim 0.2$, a signal statistical significance of 10 or higher is achievable, as shown by the yellow region.
\begin{figure}[h!]
\centering
\includegraphics[angle=0,height=7.5cm,width=8.5cm]{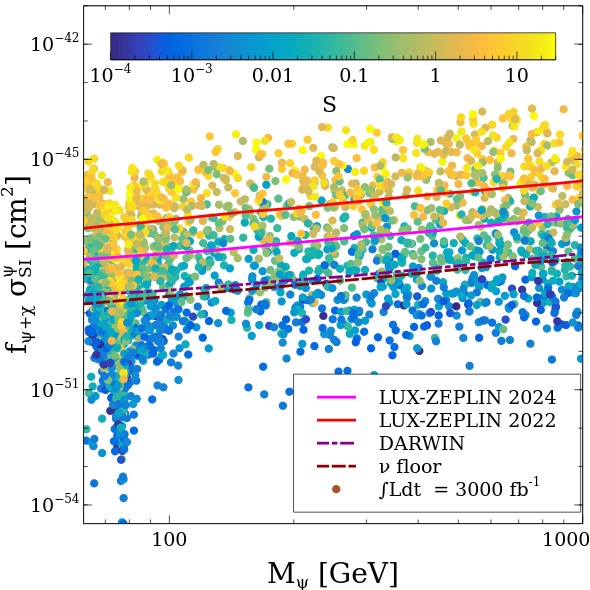}
\includegraphics[angle=0,height=7.5cm,width=8.5cm]{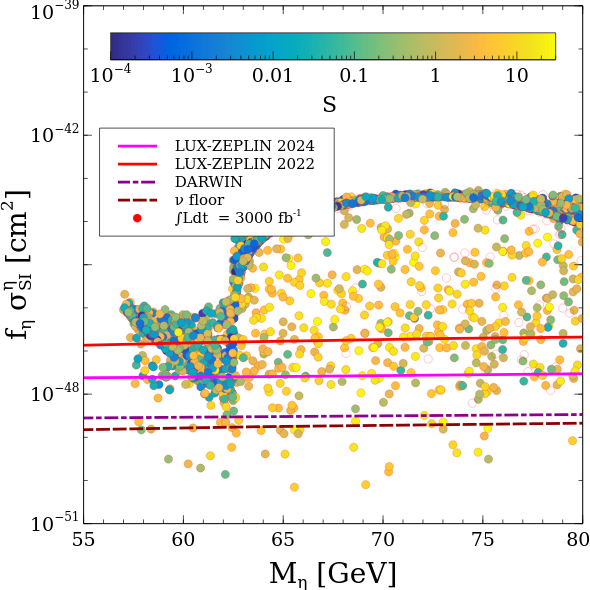}
\caption{LP (RP) shows the scatter plot in the $M_{\psi}-f_{\psi+\chi}\sigma_{SI}$  ($M_\eta -f_{\eta }\sigma_{SI}$)
plane with the colour code providing the statistical significance for the missing transverse Energy and multiple jets 
signal at LHC. The red/magenta/violet lines show present and future DD bounds, 
while the brown line gives 
as well the neutrino floor. }
\label{mh2-sigsi-S-L-1}
\end{figure}

Finally, we consider the correlation between the LHC signal and direct detection. In Fig. \ref{mh2-sigsi-S-L-1}, we show scatter plots in the $M_{\psi}-f_{\psi + \chi} \sigma_{SI}$ plane.
In the LP and RP, the colour coding indicates the statistical significance for a fixed integrated luminosity of 3000 $fb^{-1}$ for the fermionic and scalar DM, respectively.
In the LP, we see the dependence of $f_{\psi+\chi} \sigma_{SI}$ on the DM and scalar masses, where higher SIDD cross-sections generally imply 
a smaller scalar mass or a larger mixing angle, and thus a higher statistical significance for a fixed value of luminosity at the 14 TeV run of the LHC.
The dark magenta dashed-dot line in both plots corresponds to the projected sensitivity of the proposed DARWIN experiment \cite{DARWIN:2016hyl} for a 200 $t \times y$ run. The dark-red dashed line represents the $\nu$-floor, beyond which a neutrino signal should also appear in direct detection experiments, making the detection of DM more challenging. The red and magenta lines correspond to the recent bounds by the LUX-ZEPLIN experiment for different sizes of the detector volume and exposure time.
As we move to higher values of $M_{\psi}$, the statistical significance decreases for 3000 $fb^{-1}$ luminosities, and such low significance is not foreseeable in the future plans of the LHC. However, these regions can be accessed by direct detection experiments. In fact, for most of the parameter space, DD experiments could detect a signal long before collider detection of $h_{2}$. 
In the RP, we show the same analysis for the scalar DM, and there no direct correlation between the collider statistical 
significance and DD cross-section is present, 
since the direct detection cross-section is in this case not simply
suppressed
by the mixing angle as for the fermionic DM case.
So for the scalar component, the collider and DD
experiments are more complementary.
A significant portion of the scalar DM has already been excluded by the recent LUX-ZEPLIN (2024) data, 
due to the large SM Higgs mediated scattering cross-section 
and primarily the regions dominated by the beyond-SM 
Higgs resonance survive and may have a good chance to be detected at LHC.
In general both the fermionic and scalar components could appear in DD experiments soon and hopefully will be
disentangled if the DM masses are sufficiently far apart.

\section{Detection prospects at Mathusla}

In this model long-lived states like the fermions $ \psi^\prime, \chi^\prime $ can be also 
produced by the dark Higgs, and so we can actually expect to have another signal from $ h_{2} $
decay at LHC experiments and the Mathusla detector as displaced vertices. 
Indeed the lifetime of the fermions
$ \psi^\prime, \chi^\prime $ is long as it is suppressed by the small 
coupling $ \bar \alpha_j $ and
by the sterile-active neutrino mixing angle.
In this model, we generate the neutrino mass by the Type-I seesaw mechanism from 
two heavier RH neutrinos,  i.e.
\begin{eqnarray}
m_{\nu} &=& - M_{D} M_{N}^{-1} M_{D}^{T}\,, \nonumber 
\label{neutrino-mass}
\end{eqnarray}
where the Dirac and the RH neutrino mass matrix can be
expresses as,
\begin{eqnarray}
{M_{D}}_{i\,j} =\frac{y_{i\,j}\, v}{\sqrt{2}} = \frac{v}{\sqrt{2}}
\left(\begin{array}{cc}
y_{ee} ~~&~~ y_{e\mu}^{R} + i y_{e\mu}^{I} \\
y_{\mu e}~~&~~ y_{\mu \mu}^{R} + i y_{\mu \mu}^{I}\\
y_{\tau e} ~~&~~ y_{\tau \mu}^{R} + i y_{\tau \mu}^{I} 
\end{array}\right), \,\,\, 
M_{N} =\left(\begin{array}{cc}
M_{N_1} ~~&~~ 0 \\
0 ~~&~~ M_{N_2} 
\end{array}\right)\,.
\label{dirac-mass-matrix}
\end{eqnarray}
The precise measurement of neutrino oscillation data can constrain the model 
parameters, as studied in Ref. \cite{Biswas:2018sib}. 
We can easily satisfy the oscillation data \cite{Esteban:2024eli} 
even with only two RH neutrinos because there are sufficient free parameters in the
mass matrices.
The important quantity in the neutrino sector that determines the decay width of $\psi^{\prime}$ and $\chi^{\prime}$ 
is the active-sterile mixing,
given by the matrix $ M_{D} M^{-1}_{N}$, allowing to turn the intermediate heavy RH neutrino into a light SM neutrino.
The two dominant decay channels are the invisible two-body decay $\psi^\prime/\chi^\prime \rightarrow
\eta \nu$ and the three-body decay into a charged lepton and EW gauge boson
 $\psi^\prime \rightarrow \eta \ell^{\pm} W^{\mp}$.
We give the decay rates of the primed fermions in Eqs. (\ref{2-body-decay-nu}, \ref{3-body-decay}) in the
Appendix. We can easily reach decay lengths in the range that can be detected by MATHUSLA detector 
\cite{Curtin:2018mvb}. 

\begin{figure}[h!]
\centering
\includegraphics[angle=0,height=7.5cm,width=8.5cm]{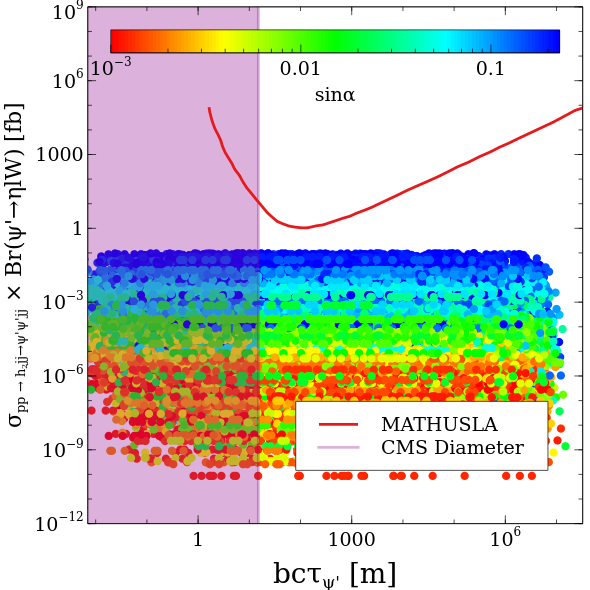}
\includegraphics[angle=0,height=7.5cm,width=8.5cm]{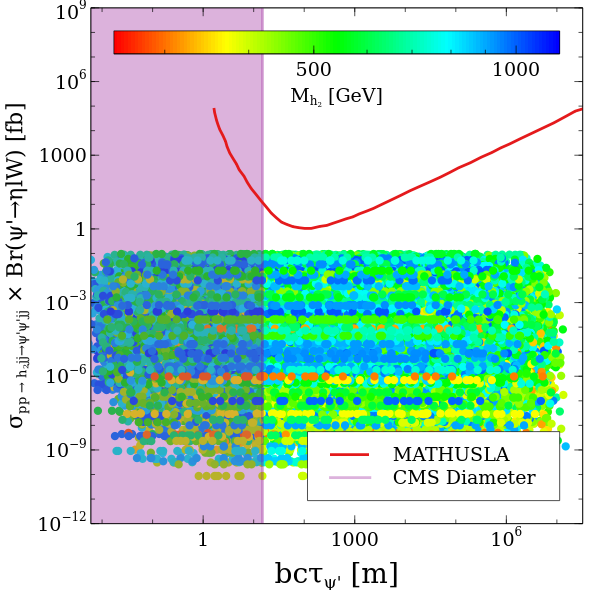}
\includegraphics[angle=0,height=7.5cm,width=8.5cm]{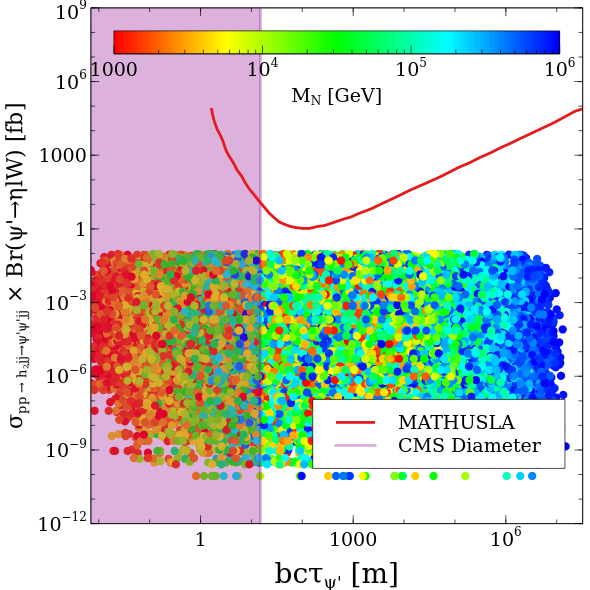}
\includegraphics[angle=0,height=7.5cm,width=8.5cm]{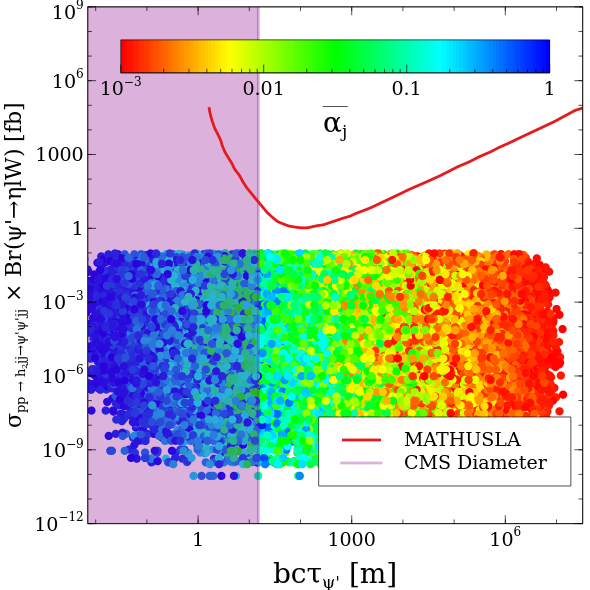}
\caption{Detection prospects of the present model at the proposed 
MATHUSLA detector. The colour bars in different plots show the different 
values of the mixing angle $\sin\alpha$, different values of BSM Higgs 
mass $M_{h_{2}}$, RH neutrino mass $M_{N}$ and the 
Yukawa coupling $\alpha_{j}$. The other parameters value and masses
have been varied as listed in Eq. \ref{range-of-parameter}.} 
\label{mathusla}
\end{figure}

The visible signal comes from the three-body decay mode, 
therefore, we consider only the fraction of metastable fermions that decay into 
a charged lepton, multiplying the production cross-section for $ \psi^\prime, \chi^\prime $ 
by the branching ratio of the three-body decay. The signal rate is then the
effective production cross-section of (at least) one lepton
\begin{eqnarray}
  \sigma_{pp \rightarrow h_{2}jj \rightarrow \psi^{\prime} \bar \psi^{\prime} jj \rightarrow \ell W j j  }
= \sigma_{pp \rightarrow h_{2} j j} \times 2\,Br(h_{2} \rightarrow \psi^{\prime} 
\bar \psi^{\prime}) \times  Br(\psi^\prime \rightarrow \eta \ell^{\pm} W^{\mp}) \,.
 \end{eqnarray} 
where the factor of $2$ comes due to two fermions 
$\psi^{\prime}$ and $\chi^{\prime}$ produce by Higgs decay. 
Note that the branching fraction is independent of the value of the couplings $ \bar \alpha_j $
and neutrino mixing, which are the same in the 2-body and 3-body decay, but depends
on the available phase space. In the limit of vanishing $ M_\eta $ there is an upper limit
on the branching fraction given as
\begin{eqnarray}
BR_{\psi^\prime \rightarrow \eta W^{\pm} \ell_i^{\mp}} &\sim& 
\frac{\Gamma_{\psi^{\prime} \rightarrow \eta W^{\pm} \ell_i^{\mp}}}{\Gamma_{\psi^\prime \rightarrow \eta \nu_i}}
\leq \frac{\alpha_{EM}}{32 \pi \; \sin^2 \theta_{W} }  \frac{M_{\psi^\prime}^2}{6 M_W^2} 
= 0.8 \times 10^{-3} \left( \frac{M_{\psi^\prime}}{300\; \mbox{GeV}} \right)^2 \; .
\label{3-body-decay-Metazero-max}
\end{eqnarray}

In Fig. \ref{mathusla}, we present the detection prospects of the metastable states at the 
proposed MATHUSLA detector \cite{Curtin:2018mvb}.  
Here we have chosen parameter points as listed in Eq. \ref{range-of-parameter} 
that allow for the 3-body decay of the metastable fermions, 
specifically 
$M_{h_{2}} > 2 M_{\psi^{\prime}}$ and $M_{\psi^{\prime}} > M_{\eta_{1}} + M_\ell + M_{W} $.  
For lower $ \psi^\prime $ masses the channel with off-shell $W$ could appear and also contribute 
to the signal, but we ignore it here, as the branching ratio is then even more suppressed.
 In the plots, the x-axis represents the total decay length 
 %  of the 3-body decay 
 % $ c \tau_{\psi^\prime} = \frac{c}{\Gamma_{\psi^{\prime} \rightarrow \eta_{1} W^{\pm} l^{\mp}}}$
 corrected with the boost factor $b$, which we take to
be $b = \frac{1.5 M_{h_{2}}}{2 M_{\psi'}}$ as described in Ref. \cite{Curtin:2018mvb}.
 
 In all plots, the red line represents the MATHUSLA projection sensitivity for the detector size [$(200 \times 200 \times 20)$ m].
 Unfortunately the rates are about one order of magnitude low for MATHUSLA sensitivity. 
 Indeed, in the region where the $ h_2 $ production cross-section is substantial, at low masses, the fermion's branching 
 ratio into lepton and $W$ is too small, while at higher masses where the branching ratio is substantial, 
 the $h_2 $ cross-section falls off. The two effects compensate each other nearly completely.

In determining the sensitivity plot in the plane of decay length and production cross section, the MATHUSLA 
collaboration has considered the decay probability distribution depending on the location of the detector 
from the interacting point using the following factor,
\begin{eqnarray}
P\left(b c \tau, L_{1}, L_{2}\right) &=& e^{-\frac{L_{1}}{b c \tau}} - e^{-\frac{L_{2}}{b c \tau}},, \nonumber \\
&\simeq& \frac{L_{2}-L_{1}}{b c \tau}\,\,\,{\rm for}\,\,\,
\left( L_{2} - L_{1} \right) \ll b c \tau,,
\end{eqnarray}
where $L_{1}$ and $L_2$  are the distances from the interaction point to the entry and 
exit of the detector, respectively. Therefore, depending on the number of long-lived 
particles produced at the interaction point,  
there may be the possibility of a signal at the MATHUSLA detector even if the long-lived 
particle decay length is larger or smaller than the detector location.

The decay length of $\psi^{\prime}$ does not depend on the mixing angle 
$\sin\alpha$ and BSM Higgs mass $M_{h_2}$, so in the top-left panel
we obtain horizontal lines  of the same colour according to the $h_2 $ production rate.
In the top-right panel instead the horizontal pattern is less definite, since the 
$ h_2 $ mass affects the production rate, the $ h_2 $ branching fraction in
the primed fermions and the boost factor $b$.
In the lower two plots, we show the dependence on the
coupling with RH neutrinos $ \bar \alpha_j $ and the RH neutrino
mass, which affect  $\psi^{\prime}$ DM decay length. Here we keep the
light neutrino masses fixed to scale of the 
atmospheric mass difference.
The production rate of $\psi^{\prime}$ does not depend on these
parameters, therefore we see a vertical pattern in the colors. 
Note that we can neither take $\bar \alpha_{j}$ 
arbitrarily small nor $M_{N}$ arbitrarily 
large because then $\psi^\prime $ decays during or
after BBN and may change its successful predictions. 
 The Yukawa coupling $\bar \alpha_{j}$ and heavy neutrino mass $ M_N$ 
 can be constrained  by demanding 
 $\tau_{\psi^{\prime}} < 1$ sec as follows,
\begin{eqnarray}
\bar \alpha_{j} \geq 1.82 \times 10^{-3} 
\left(\frac{10^{-11}}{m_{\nu}} \right)^{1/2}
\left(\frac{M_{N}}{10^{6}\,\,{\rm GeV}} \right)^{1/2}
\left(\frac{1 \,\,{\rm GeV}}{M_{\psi^{\prime}}} \right)^{1/2}\,.
\end{eqnarray}

In all the plots, we show the CMS detector size as the magenta-shaded region.
Thus, we see that some of the metastable fermions could give rise to displaced vertices 
and leptons within the LHC detectors.
The CMS and ATLAS collaborations have performed studies for Long-Lived-Particles
(LLPs). In particular, CMS has completed searches in models where the SM Higgs mainly 
decays to LLP particles, which then decay to BSM particles, already placing bounds on 
decay lengths between $10^{-5}$ m  and $10^{4}$ m \cite{CMS:2024qxz}, 
which covers the area of our interest in the present work. 
In particular, the channels considered were SM Higgs decays to dark gauge bosons 
and to s-quarks \cite{CMS:2024qxz}, as well as decays to scalars 
\cite{CMS:2021juv}. Long-lived heavy neutral lepton have been also searched
for \cite{CMS:2024hik}. 
Unfortunately none of those studies is directly applicable to our model,
since either the production rate or the final state are different.
Possibly the most promise channel could be the primed fermion decays into
muons that can be observed in the muon chamber far away from the central
region, as considered in  \cite{CMS:2024qxz}, but without imposing the
constraint that the dimuon system originates from the same particle.
Since only a fraction of the $h_2$ decays into metastable fermions and
only a fraction of them decays within the CMS detector, we expect that
the statistics will play in favour of a first detection of $h_2$ via the missing energy
signal, but the search for displaced vertices could be used later to give
more information on the model.

We can see that at the 14 TeV run of LHC, it is difficult to observe this
particular scenario via LLPs because of small production rate and 
small branching fraction of exotic fermion decay into charged lepton. 

The parameter space can become accessible at the proposed 100 TeV pp collider because 
there the SM Higgs boson production rate will be significantly enhanced, almost by a 
factor of $\mathcal{O}(20)$ \cite{Baglio:2015wcg}.  
Moreover, the 100 TeV pp collider will also increase the possibility to access more 
massive $\psi^{\prime}$, and we have checked that higher mass of $\psi^{\prime}$ can 
lead to branching as large as $\mathcal{O}(10^{-2})$.

%%%%
\begin{figure}[h!]
\centering
\includegraphics[angle=0,height=7.5cm,width=8.5cm]{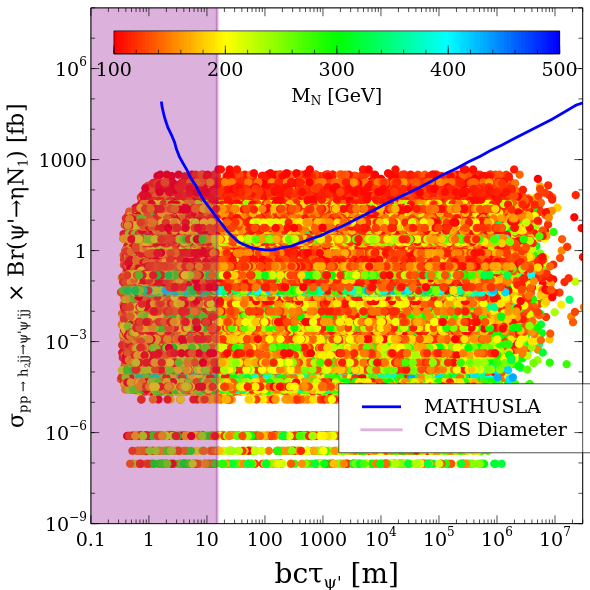}
\includegraphics[angle=0,height=7.5cm,width=8.5cm]{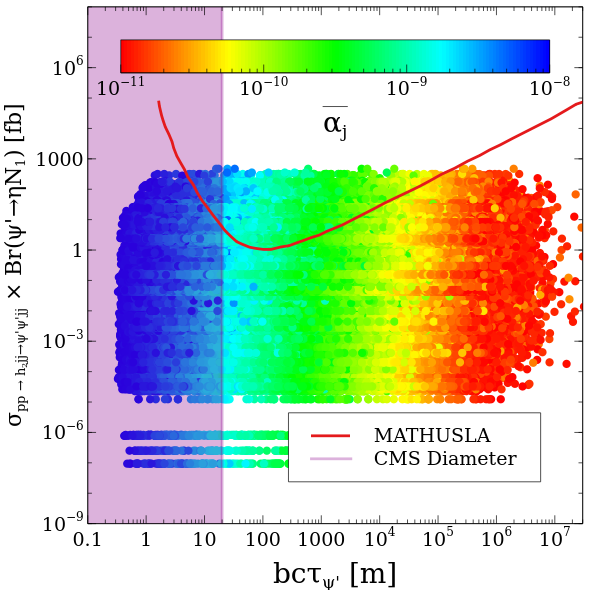}
\caption{Detection prospects of the present model at the proposed 
MATHUSLA detector. The colour bars in different plots show the different 
values of the RH neutrino mass $M_{N}$ and the 
Yukawa coupling $\bar\alpha_{j}$.} 
\label{mathusla-2}
\end{figure}
%%%

Another interesting range of the parameter space to explore is where the $\psi^{\prime}$ can 
directly decay to on-shell RH neutrino $N_{i}$ and $\eta $ for lower values of RH neutrino mass 
$M_{N_i} \sim \mathcal{O}(100)$ GeV. 
The produced RH neutrino will then decay to the SM particles $\ell^{\mp}W^{\pm}, \ell^{\mp} \ell^{\pm} \nu, \ell jj$ at the MATHUSLA detector.  
Assuming that the branching ratio of $\psi^{\prime} \rightarrow N \eta_i$ is $100\%$, in Fig. \ref{mathusla-2}, 
we show this scenario and we can see that it can be accessed at the MATHUSLA
detector for a wide range of decay lengths if the Yukawa coupling associated with the RH
neutrino and dark sector is $\bar \alpha_{j} \sim \mathcal{O}(10^{-9})$. 
In the plot, we keep the RH neutrino mass in the range $100 \leq M_{N_1}\,\,{\rm GeV}\leq 500$ GeV and $10^{-11} \leq \bar \alpha_{j} \leq 10^{-8}$
as shown in the colour bar.
We can see from the LP that the low mass of the RH neutrino will be accessible at MATHUSLA 
while higher masses are difficult to be produced.  To summarise, we find that the proposed MATHUSLA 
detector can observe the metastable fermions for small RH neutrino mass and small couplings $\bar \alpha_{j}$. 
Similar kinds of detection prospects can also be expected at the FASER detector \cite{FASER:2022hcn}.
    
As a concluding comment, in the asymmetric DM scenario the $ \psi^\prime, \chi^\prime $ are stable, but in this
case a signal at MATHUSLA could still arise from the decay of the metastable $ \eta $ states with similar decay rates 
since the decay is determined by the same parameters $ \bar \alpha_j $ and $ M_R $. The production rate 
can take place again from $ h_2 $ decay and be comparable, as long as $ \eta $  is not too heavy. 
Note though that, for vanilla leptogenesis and asymmetric DM the RH neutrino mass needs to be above 
$ 10^8 $ GeV and therefore the neutrino mixing will become too small to fall in the MATHUSLA range. 
Therefore the appearance of a signal would be possible only for a non-standard leptogenesis scenario. 
%%%%%%%%%%%%%%%%%%%%%%%%%%%%%%%%%%%

\section{Summary and Conclusion}
\label{conclusion}

In this work, we have extended the study of one simple model explaining 
dark matter, neutrino mass and matter anti-matter asymmetry of the Universe
in a minimal setting.  
In the dark sector we impose an $SU(2)_D $ gauge symmetry
and two discrete symmetries leading in general to two stable particles.
Indeed, the model contains a plethora of Dark Matter candidates
and can accommodate both symmetric or antisymmetric Dark Matter. 
Since the case of asymmetric DM has already been considered in \cite{Biswas:2018sib}, 
we concentrate here on the case of symmetric dark matter and on the 
collider detection prospects of the model connected to the extended Higgs 
sector.

To explore the main qualitative features of the model with a WIMP DM, 
we consider degenerate masses in the fermionic and bosonic
sectors so that effectively we reduce the system to two coupled Boltzmann
equations for the relic densities of the scalar and fermionic doublets.
Moreover, we fix the dark gauge boson mass between the fermionic DM 
mass and the scalar doublet mass so that the gauge bosons decays 
quickly into the exotic scalars.
We have found that the fermionic DM candidates mostly annihilate 
into dark gauge bosons and therefore have a small relic density in most
parameter space. On the other hand, the dark scalar can be overproduced
via the mixing with the SM doublet Higgs and in the light mass region only 
a resonant annihilation it can provide the correct DM density.
We concentrate on the scalar mass at the SM Higgs resonance so
that the dark gauge boson and the second Higgs boson can both
decay into the dark scalar doublet.

In the model both the fermionic and scalar Dark Matter candidates couple
to the SM Higgs bosons via the mixing and therefore can be 
detected by Higgs mediated elastic scattering in direct detection experiments.
A large region for scalar doublet masses above half of the SM Higgs mass are
ruled out by direct detection and the lower mass region will be accessed in future. 
Also for the fermionic DM, the parameter region with a large 
mixing angle is already ruled out by the  LUX-ZEPLIN DD experiment and the 
rest of the allowed region will be explored in the future. 
Even if the fermions give a subdominant contribution to the relic density for 
$g_{D} \sim \mathcal{O}(1)$,  the signal in direct detection can still
be visible and even stronger than that of the dominant scalar component.
Note though that there is no lower bound on the DD signal since for both DM 
particles a partial cancellation can reduce the scattering cross-section 
and for the fermionic DM also the relic density can be strongly suppressed for 
large, but reasonable values of $g_D$.

Regarding indirect detection, both the scalar and fermionic DM components 
are below the present Fermi-LAT bound, if the scalar mass is not exactly
at the Higgs resonance. In this case, the scalar component gives a larger signal
 thanks to the larger cross-section and density.

We have explored the prospect to detect the model at LHC
via the production of the BSM Higgs $ h_{2} $ decaying invisibly in the
fermionic Dark Matter. The signal expected in this case is missing
$ E_T $ with one or more jets as recoiling particles. The
same signal can also appear in the scenario with asymmetric Dark Matter
as the Higgs mixing and the BSM Higgs decay channels can be the same.
We have analysed the $\cancel{E}_{T} + 2j$ signal at the 14 TeV run of the pp 
collider performing simulations for both signal and background and
devised optimised cuts to reduce the overwhelming SM background.
With appropriate cuts, it is possible to extract the BSM Higgs signal, 
especially in the low-mass region. 
A  BSM Higgs with mixing angle $ \sin\alpha = 0.2 $ could be observed at
LHC well above the 5 $\sigma $ significance, and even a 600 GeV mass
could be reached at 5 $\sigma $ level at the end of the high luminosity phase.
For larger masses, $ h_2 $ production at colliders becomes more difficult, 
but direct detection can instead cover the parameter space at large mixing.
The two searches are complementary and both detections could help
in excluding the large mixing angle region for any BSM Higgs mass.

Finally, in the model we have also metastable fermions that
can be produced from the BSM Higgs; we have studied the
prospect of their detection in the searches for long-lived particles
and in particular in the MATHUSLA detector. 
Since the lifetime of the metastable dark fermions is determined by
the neutrino see-saw parameters and in particular the RH neutrino mass,
they are naturally very long-lived. Since only the subleading 3-body 
decay produces visible particles, though, only a fraction of the
metastable fermions is visible at MATHUSLA. 
We find that at the 14 TeV run of LHC the rates are too small to explore the 
allowed parameter space and one needs to consider a 100 TeV collider
to reach a rate measurable by the MATHUSLA detector. 
Some of the metastable particles may decay even within the LHC detectors,
giving displaced vertices at meter distances from the interaction point and
could maybe be detected if the experimental analyses are extended.
Moreover, if the mass of the right handed neutrino is around 100 GeV and 
$\bar \alpha_j \sim 10^{-9}$, the exotic fermions can decay into scalar and
RH neutrino and this gives another possible signal at the MATHUSLA 
detector.

All in all, our model is very rich and provides interesting signals
at many different experiments, and a combined analysis of all
the signals will probably be needed to identify and reconstruct the model
and identify the Dark Matter type and its production mechanism.  
Most parameter space with a dominant scalar WIMP DM will be covered 
by the next generation of direct detection experiments, while the case with mainly 
an asymmetric fermionic DM component is more difficult to see due to
the Higgs mixing angle suppression.
The search for the scalar particles mixing with the SM Higgs is in any case 
the best opportunity of discovery for this type of models and we encourage
the experimental collaboration to pursue these searches in the future.

%%%%%%%%%%%%%%%%%%%%%%%%%%%%%%%%%%%%%%%%%%%%%%%%%%%%%%%%%%%%%%%
\section{Acknowledgements}

SK expresses gratitude for the warm hospitality extended at HRI, Prayagraj, where a 
portion of the work was undertaken.  
This project has received funding from the European Union's Horizon 2020 research and innovation programme 
under the Marie  Sklodowska -Curie grant agreement No 860881-HIDDeN and under the Marie Sklodowska-Curie 
Staff Exchange grant agreement No 101086085 - ASYMMETRY.
This research was supported by Brain Pool program funded by the Ministry of Science and ICT through the 
National Research Foundation of Korea\,(RS-2024-00407977).
This work used the Scientific Compute Cluster at GWDG, the joint data center 
of Max Planck Society for the Advancement of Science (MPG) and 
University of G\"{o}ttingen. SD and SKR would like to acknowledge the support from the Department of Atomic Energy
(DAE), India, for the Regional Centre for Accelerator-based Particle Physics (RECAPP), Harish
Chandra Research Institute. 

%%%%%%%%%%%%%%%%%%%%%%%%%%%%%%%%%%%%%%%%%%%%%%%%%%%%%%%%%%%%%%%

\appendix
\section{Decay rates}
\label{appendix}

\subsection{\bf Decay width of $h_{2}$:}
\label{H2-decay-expression}
The Higgs $ h_{2} $  decays to the SM particles as well as into the dark sectors. 
We list here the dominant decay modes of BSM Higgs boson mainly to the 
dark sector and the associated vertices, while the decay modes in the SM
particles are similar to the SM Higgs ones, apart for the mixing angle.

\begin{itemize}
\item {\bf $h_{2} \rightarrow W_{D} W_{D}$:} The decay width of
$h_{2}$ to the dark gauge bosons are given as,
\begin{eqnarray}
\Gamma_{h_{2} \rightarrow W_D W_D} = 
\frac{M^3_{h_{2}} g^{2}_{h_{2}W_D W_D}}{128 \pi M^{4}_{W_{D}}} \sqrt{1 - \frac{4 M^2_{W_D}}{M^2_{h_2}}} 
\left( 1 - \frac{4 M^2_{W_D}}{M^2_{h_2}} + \frac{12 M^4_{W_D}}{M^4_{h_2}} \right)
\label{H2-WiWi}
\end{eqnarray}
where $g_{h_{2}W_{D}W_{D}} = \cos\alpha\, g_{D} M_{W_D}$.

\item {\bf $h_{2} \rightarrow f \bar{f}$:} The decay width for the BSM
Higgs decay to the SM/BSM fermions can be expressed as,
\begin{eqnarray}
\Gamma_{h_{2} \rightarrow f \bar{f}} = \frac{n_{c} M_{h_2} g^2_{h_{2}f \bar{f}}}{8 \pi} \left(1 -\frac{4 M^2_{f}}{M^2_{h_2}} \right)^{3/2}\,,
\end{eqnarray}
where $n_{c}$ is number of colors and the $h_{2}$ coupling to the 
fermions are given as,
\begin{eqnarray}
g_{h_{2}f \bar{f}} &=&  \frac{\sin\alpha M_{f}}{v} \,\,\,{\rm for\,\,\,SM \,\,\, fermions}\,, \nonumber \\
&=& \frac{\cos\alpha M_{f}}{v_{D}} \,\,\,{\rm for\,\,\,BSM \,\,\, fermions}\,.
\label{H2-psi-psi}
\end{eqnarray}

\item {\bf $h_{2} \rightarrow \eta \bar \eta$:} The decay width of BSM Higgs to dark 
scalars take the following form,
\begin{eqnarray}
\Gamma_{h_{2} \rightarrow \eta_i \eta_i^{\dagger}} = \frac{g^{2}_{h_{2}\eta_{i} \eta_{i}}}{32 \pi M_{h_{2}} }
\sqrt{1 - \frac{4 M^2_{\eta_{i}}}{M^2_{h_2}}}
\label{H2-ss}
\end{eqnarray}
where the coupling 
$g_{h_{2}\eta_{i}\eta_{i}}$ has been defined in Eq. (\ref{h-eta-eta}) and contains the two v.e.v.s.

\subsection{\bf Decay Width of $\psi^\prime, \chi^\prime $:}

{\bf Two body decay:}

The next-to-lightest stable particles, $\psi^{\prime}$ and $\chi^{\prime}$, 
can decay to a light neutrino and $\eta_{1,2}$. The decay rate can be 
expressed as,
\begin{eqnarray}
\Gamma_{\psi^{\prime} \rightarrow \nu_i \eta_{j}} &=&  
\frac{ \sum_j  (\theta^2_{\nu N})_{ij} \bar \alpha^2_{j} M_{\psi^{\prime}}}{16 \pi } \left( 1 - \frac{M^2_{\eta}}{M_{\psi^{\prime}}^2} - \frac{m^2_{\nu}}{M_{\psi^{\prime}}^2}  \right)
\sqrt{1 - \left( \frac{M_{\eta} + m_{\nu}}{M_{\psi^{\prime}}} \right)^{2}}
\sqrt{1 - \left( \frac{M_{\eta} - m_{\nu}}{M_{\psi^\prime}} \right)^{2}}
\nonumber\\
&=&  
 \frac{ \sum_j  (\theta^2_{\nu N})_{ij} \bar \alpha^2_{j} M_{\psi^\prime}}{16 \pi} 
 \left( 1 -   \frac{M_\eta^2 }{M_{\psi^\prime}^2}  \right)^2
\label{2-body-decay-nu}
\end{eqnarray}
where $\eta_{j} = \eta_{1,2}$, $ (\theta_{\nu N})_{ij} = (M_{D}M^{-1}_{R})_{ij}$ ( $ i = 1,2,3 , j=1, 2 $) is the
active sterile mixing matrix, $ \bar \alpha_j $ is the coupling of RH neutrino to the dark sector 
 and in the last line, we have neglected the neutrino mass.

Similarly, the decay length for
 the $\psi^{\prime} \rightarrow N_{j} \eta_{j}$ can be expressed as,
 \begin{eqnarray}
\Gamma_{\psi^{\prime} \rightarrow N_j \eta_{i}} &=&  
\frac{   \bar \alpha^2_{j} M_{\psi^{\prime}}}{16 \pi} \left( 1 - \frac{M^2_{\eta}}{M_{\psi^{\prime}}^2} - \frac{m^2_{N_j}}{M_{\psi^{\prime}}^2}  \right)
\sqrt{1 - \left( \frac{M_{\eta} + m_{N_j}}{M_{\psi^{\prime}}} \right)^{2}}
\sqrt{1 - \left( \frac{M_{\eta} - m_{N_j}}{M_{\psi^\prime}} \right)^{2}}
\nonumber\\
\label{2-body-decay-N}
\end{eqnarray}

{\bf Three body decay:}

A subleading decay of  $\psi^{\prime}$ and $\chi^{\prime}$ can produce visible charged particles, 
in particular $  \ell W $, emitted from the light neutrino. So this decay channel is suppressed
by the small coupling $ \bar \alpha_j $ and the neutrino mixing angle as in Eq. (\ref{2-body-decay-nu}).
The three-body decay width for $\psi^{\prime}, \chi^{\prime} \rightarrow \eta_{1,2} W^{\pm} \ell^{\mp}$ 
can be expressed as,
\begin{eqnarray}
\Gamma_{\psi^{\prime} \rightarrow \eta W^{\pm} \ell_i^{\mp}} &=& 
\frac{\sum_j  (\theta^2_{\nu N})_{ij} \bar \alpha^2_{j}  \alpha_{EM} \pi }{2 \sin^2 \theta_{W} } 
\frac{M_{\psi^{\prime}}}{32 (2\pi)^{3}}  
\int_{x^{-}_{3}}^{x^{+}_{3}} \int_{x^{-}_{1}}^{x^{+}_{1}} | {\cal M} |^{2} d x_{1} d x_{3}\label{3-body-decay}
\end{eqnarray}
where, neglecting the lepton mass,
\begin{align}
  |{\cal M} |^2 =&  
 \frac{1}{2 a_2  \left(1 - x_1 + a_1) \right)^2}
\biggl( 
 (1- x_1 )^2 x_3 + 2 a_2 (1-x_1)(2-x_1-x_3) + 2 a_1 (1-x_1) x_3
  \nonumber \\&
 - 2 a_2^2 (2 - x_1) 
+ 2 a_1 a_2  (2- x_1 - x_3)
 + a_1^2 x_3
   \biggr) 
 \biggr)\,, \nonumber \\
 x^{\pm}_{1} =& 1 + a_{1} - a_{2} - x_{3} + \frac{x_{3}}{2} 
  \left( 1 + \frac{a_{1} - a_{2}}{1 -x_{3}} \right) \pm \frac{x_{3} }{2}
\times 
 \sqrt{1-2 \frac{a_{1} + a_{2}}{1 - x_{3} } + \frac{(a_{1}-a_{2})^{2}}
{(1 -x_{3})^{2}} }\,\,,\,\nonumber \\  
 x^{-}_{3} =&\,\,  0,\,  x^{+}_{3} =  1  - a_{1} - a_{2} - 2 \sqrt{a_{1} a_{2}} \,\,\,
 \text{and}\,\,\, a_{1} = \left( \frac{M_{\eta}}{M_{\psi^{\prime}}} \right)^{2}, 
a_{2} = \left( \frac{M_{W}}{M_{\psi^{\prime}}} \right)^{2}.
\end{align}
In the limit of vanishing $ M_\eta $ the expressions simplify and the
3-body decay rate can be given as an analytic formula as
\begin{eqnarray}
\Gamma_{\psi^{\prime} \rightarrow \eta W^{\pm} \ell_i^{\mp}} &=& 
\frac{\alpha_{EM}  \Gamma_{\psi^\prime \rightarrow \eta \nu_i} }{32 \pi \; \sin^2 \theta_{W} } 
 \int_{0}^{1-a_2}\!\!\!  d x_3  \int_{1-x_3-a_2}^{1- \frac{a_2}{1-x_3}}\!\!\! d x_1 
 \left[\frac{x_3}{2 a_2} + 1 + \frac{1-x_3-a_2}{1-x_1} - \frac{a_2}{(1-x_1)^2}  \right] 
 \nonumber \\
&=& 
\Gamma_{\psi^\prime \rightarrow \eta \nu_i} 
\frac{\alpha_{EM}}{32 \pi \; \sin^2 \theta_{W} } \frac{1}{6 a_2}
\left[1- \frac{9}{2} a_2 - 9 a_2^2 - \frac{11}{2} a_2^3 + 3 a_2^3 \log (a_2) \right]
\label{3-body-decay-Metazero}
\end{eqnarray}
where the expression in the square bracket can be maximally one for $ a_2 = 0 $.

\end{itemize}


\begin{thebibliography}{99}

%\cite{Sofue:2000jx}
\bibitem{Sofue:2000jx}
Y.~Sofue and V.~Rubin,
%``Rotation curves of spiral galaxies,''
Ann. Rev. Astron. Astrophys. \textbf{39}, 137-174 (2001)
doi:10.1146/annurev.astro.39.1.137
[arXiv:astro-ph/0010594 [astro-ph]].
%794 citations counted in INSPIRE as of 26 Dec 2024

%\cite{Clowe:2003tk}
\bibitem{Clowe:2003tk}
D.~Clowe, A.~Gonzalez and M.~Markevitch,
%``Weak lensing mass reconstruction of the interacting cluster 1E0657-558: Direct evidence for the existence of dark matter,''
Astrophys. J. \textbf{604}, 596-603 (2004)
doi:10.1086/381970
[arXiv:astro-ph/0312273 [astro-ph]].
%743 citations counted in INSPIRE as of 20 Dec 2024

%\cite{Planck:2018vyg}
\bibitem{Planck:2018vyg}
N.~Aghanim \textit{et al.} [Planck],
%``Planck 2018 results. VI. Cosmological parameters,''
Astron. Astrophys. \textbf{641}, A6 (2020)
[erratum: Astron. Astrophys. \textbf{652}, C4 (2021)]
doi:10.1051/0004-6361/201833910
[arXiv:1807.06209 [astro-ph.CO]].
%15943 citations counted in INSPIRE as of 29 Dec 2024

%\cite{Fields:2014uja}
\bibitem{Fields:2014uja}
B.~D.~Fields, P.~Molaro and S.~Sarkar,
%``Big-Bang Nucleosynthesis,''
Chin. Phys. C \textbf{38}, 339-344 (2014)
[arXiv:1412.1408 [astro-ph.CO]].
%105 citations counted in INSPIRE as of 02 Dec 2024

%\cite{Super-Kamiokande:1998kpq}
\bibitem{Super-Kamiokande:1998kpq}
Y.~Fukuda \textit{et al.} [Super-Kamiokande],
%``Evidence for oscillation of atmospheric neutrinos,''
Phys. Rev. Lett. \textbf{81}, 1562-1567 (1998)
doi:10.1103/PhysRevLett.81.1562
[arXiv:hep-ex/9807003 [hep-ex]].
%8308 citations counted in INSPIRE as of 29 Dec 2024

%\cite{SNO:2002tuh}
\bibitem{SNO:2002tuh}
Q.~R.~Ahmad \textit{et al.} [SNO],
%``Direct evidence for neutrino flavor transformation from neutral current interactions in the Sudbury Neutrino Observatory,''
Phys. Rev. Lett. \textbf{89}, 011301 (2002)
doi:10.1103/PhysRevLett.89.011301
[arXiv:nucl-ex/0204008 [nucl-ex]].
%4757 citations counted in INSPIRE as of 24 Dec 2024

%\cite{Biswas:2018sib}
\bibitem{Biswas:2018sib}
A.~Biswas, S.~Choubey, L.~Covi and S.~Khan,
%``Common origin of baryon asymmetry, dark matter and neutrino mass,''
JHEP \textbf{05}, 193 (2019)
doi:10.1007/JHEP05(2019)193
[arXiv:1812.06122 [hep-ph]].
%15 citations counted in INSPIRE as of 22 Nov 2024

%\cite{ATLAS:2017uhp}
\bibitem{ATLAS:2017uhp}
M.~Aaboud \textit{et al.} [ATLAS],
%``Search for heavy resonances decaying into $WW$ in the $e\nu\mu\nu$ final state in $pp$ collisions at $\sqrt{s}=13$ TeV with the ATLAS detector,''
Eur. Phys. J. C \textbf{78}, no.1, 24 (2018)
doi:10.1140/epjc/s10052-017-5491-4
[arXiv:1710.01123 [hep-ex]].
%188 citations counted in INSPIRE as of 29 Dec 2024

%\cite{ATLAS:2017tlw}
\bibitem{ATLAS:2017tlw}
M.~Aaboud \textit{et al.} [ATLAS],
%``Search for heavy ZZ resonances in the $\ell ^+\ell ^-\ell ^+\ell ^-$ and $\ell ^+\ell ^-\nu \bar{\nu }$ final states using proton\textendash{}proton collisions at $\sqrt{s}= 13$   $\text {TeV}$ with the ATLAS detector,''
Eur. Phys. J. C \textbf{78}, no.4, 293 (2018)
doi:10.1140/epjc/s10052-018-5686-3
[arXiv:1712.06386 [hep-ex]].
%161 citations counted in INSPIRE as of 29 Dec 2024

%\cite{ATLAS:2017ayi}
\bibitem{ATLAS:2017ayi}
M.~Aaboud \textit{et al.} [ATLAS],
%``Search for new phenomena in high-mass diphoton final states using 37 fb$^{-1}$ of proton--proton collisions collected at $\sqrt{s}=13$ TeV with the ATLAS detector,''
Phys. Lett. B \textbf{775}, 105-125 (2017)
doi:10.1016/j.physletb.2017.10.039
[arXiv:1707.04147 [hep-ex]].
%230 citations counted in INSPIRE as of 29 Dec 2024

%\cite{Belanger:2021slj}
\bibitem{Belanger:2021slj}
G.~B\'elanger, S.~Khan, R.~Padhan, M.~Mitra and S.~Shil,
%``Right handed neutrinos, TeV scale BSM neutral Higgs boson, and FIMP dark matter in an EFT framework,''
Phys. Rev. D \textbf{104}, no.5, 055047 (2021)
doi:10.1103/PhysRevD.104.055047
[arXiv:2104.04373 [hep-ph]].
%11 citations counted in INSPIRE as of 29 Dec 2024

%\cite{Banerjee:2015hoa}
\bibitem{Banerjee:2015hoa}
S.~Banerjee, M.~Mitra and M.~Spannowsky,
%``Searching for a Heavy Higgs boson in a Higgs-portal B-L Model,''
Phys. Rev. D \textbf{92}, no.5, 055013 (2015)
doi:10.1103/PhysRevD.92.055013
[arXiv:1506.06415 [hep-ph]].
%22 citations counted in INSPIRE as of 29 Dec 2024

%\cite{Banerjee:2016foh}
\bibitem{Banerjee:2016foh}
S.~Banerjee, B.~Bhattacherjee, M.~Mitra and M.~Spannowsky,
%``The Lepton Flavour Violating Higgs Decays at the HL-LHC and the ILC,''
JHEP \textbf{07}, 059 (2016)
doi:10.1007/JHEP07(2016)059
[arXiv:1603.05952 [hep-ph]].
%47 citations counted in INSPIRE as of 29 Dec 2024



%\cite{Curtin:2018mvb}
\bibitem{Curtin:2018mvb}
D.~Curtin, M.~Drewes, M.~McCullough, P.~Meade, R.~N.~Mohapatra, J.~Shelton, B.~Shuve, E.~Accomando, C.~Alpigiani and S.~Antusch, \textit{et al.}
%``Long-Lived Particles at the Energy Frontier: The MATHUSLA Physics Case,''
Rept. Prog. Phys. \textbf{82}, no.11, 116201 (2019)
doi:10.1088/1361-6633/ab28d6
[arXiv:1806.07396 [hep-ph]].
%451 citations counted in INSPIRE as of 20 Dec 2024

%\cite{FASER:2022hcn}
\bibitem{FASER:2022hcn}
H.~Abreu \textit{et al.} [FASER],
%``The FASER detector,''
JINST \textbf{19}, no.05, P05066 (2024)
doi:10.1088/1748-0221/19/05/P05066
[arXiv:2207.11427 [physics.ins-det]].
%77 citations counted in INSPIRE as of 26 Dec 2024

%\cite{Bechtle:2015pma}
\bibitem{Bechtle:2015pma}
P.~Bechtle, S.~Heinemeyer, O.~Stal, T.~Stefaniak and G.~Weiglein,
%``Applying Exclusion Likelihoods from LHC Searches to Extended Higgs Sectors,''
Eur. Phys. J. C \textbf{75}, no.9, 421 (2015)
doi:10.1140/epjc/s10052-015-3650-z
[arXiv:1507.06706 [hep-ph]].
%292 citations counted in INSPIRE as of 24 Dec 2024

%\cite{Bechtle:2013xfa}
\bibitem{Bechtle:2013xfa}
P.~Bechtle, S.~Heinemeyer, O.~St\r{a}l, T.~Stefaniak and G.~Weiglein,
%``$HiggsSignals$: Confronting arbitrary Higgs sectors with measurements at the Tevatron and the LHC,''
Eur. Phys. J. C \textbf{74}, no.2, 2711 (2014)
doi:10.1140/epjc/s10052-013-2711-4
[arXiv:1305.1933 [hep-ph]].
%630 citations counted in INSPIRE as of 24 Dec 2024

%\cite{Belanger:2001fz}
\bibitem{Belanger:2001fz}
G.~Belanger, F.~Boudjema, A.~Pukhov and A.~Semenov,
%``MicrOMEGAs: A Program for calculating the relic density in the MSSM,''
Comput. Phys. Commun. \textbf{149}, 103-120 (2002)
doi:10.1016/S0010-4655(02)00596-9
[arXiv:hep-ph/0112278 [hep-ph]].
%740 citations counted in INSPIRE as of 24 Dec 2024

%\cite{Alloul:2013bka}
\bibitem{Alloul:2013bka}
A.~Alloul, N.~D.~Christensen, C.~Degrande, C.~Duhr and B.~Fuks,
%``FeynRules  2.0 - A complete toolbox for tree-level phenomenology,''
Comput. Phys. Commun. \textbf{185}, 2250-2300 (2014)
doi:10.1016/j.cpc.2014.04.012
[arXiv:1310.1921 [hep-ph]].
%2607 citations counted in INSPIRE as of 27 Dec 2024

%\cite{Belyaev:2012qa}
\bibitem{Belyaev:2012qa}
A.~Belyaev, N.~D.~Christensen and A.~Pukhov,
%``CalcHEP 3.4 for collider physics within and beyond the Standard Model,''
Comput. Phys. Commun. \textbf{184}, 1729-1769 (2013)
doi:10.1016/j.cpc.2013.01.014
[arXiv:1207.6082 [hep-ph]].
%1054 citations counted in INSPIRE as of 18 Dec 2024

%\cite{Alwall:2011uj}
\bibitem{Alwall:2011uj}
J.~Alwall, M.~Herquet, F.~Maltoni, O.~Mattelaer and T.~Stelzer,
%``MadGraph 5 : Going Beyond,''
JHEP \textbf{06}, 128 (2011)
doi:10.1007/JHEP06(2011)128
[arXiv:1106.0522 [hep-ph]].
%3956 citations counted in INSPIRE as of 19 Dec 2024

%\cite{LZ:2022lsv}
\bibitem{LZ:2022lsv}
J.~Aalbers \textit{et al.} [LZ],
%``First Dark Matter Search Results from the LUX-ZEPLIN (LZ) Experiment,''
Phys. Rev. Lett. \textbf{131}, no.4, 041002 (2023)
doi:10.1103/PhysRevLett.131.041002
[arXiv:2207.03764 [hep-ex]].
%816 citations counted in INSPIRE as of 24 Dec 2024

%\cite{LZCollaboration:2024lux}
\bibitem{LZCollaboration:2024lux}
J.~Aalbers \textit{et al.} [LZ Collaboration],
%``Dark Matter Search Results from 4.2 Tonne-Years of Exposure of the LUX-ZEPLIN (LZ) Experiment,''
[arXiv:2410.17036 [hep-ex]].
%25 citations counted in INSPIRE as of 24 Dec 2024

%\cite{MAGIC:2016xys}
\bibitem{MAGIC:2016xys}
M.~L.~Ahnen \textit{et al.} [MAGIC and Fermi-LAT],
%``Limits to Dark Matter Annihilation Cross-Section from a Combined Analysis of MAGIC and Fermi-LAT Observations of Dwarf Satellite Galaxies,''
JCAP \textbf{02}, 039 (2016)
doi:10.1088/1475-7516/2016/02/039
[arXiv:1601.06590 [astro-ph.HE]].
%474 citations counted in INSPIRE as of 24 Dec 2024

%\cite{ATLAS:2020tlo}
\bibitem{ATLAS:2020tlo}
G.~Aad \textit{et al.} [ATLAS],
%``Search for heavy resonances decaying into a pair of Z bosons in the $\ell ^+\ell ^-\ell '^+\ell '^-$ and $\ell ^+\ell ^-\nu {{\bar{\nu }}}$ final states using 139 $\mathrm {fb}^{-1}$ of proton\textendash{}proton collisions at $\sqrt{s} = 13\,$TeV with the ATLAS detector,''
Eur. Phys. J. C \textbf{81}, no.4, 332 (2021)
doi:10.1140/epjc/s10052-021-09013-y
[arXiv:2009.14791 [hep-ex]].
%143 citations counted in INSPIRE as of 23 Dec 2024

%\cite{Alwall:2014hca}
\bibitem{Alwall:2014hca}
J.~Alwall, R.~Frederix, S.~Frixione, V.~Hirschi, F.~Maltoni, O.~Mattelaer, H.~S.~Shao, T.~Stelzer, P.~Torrielli and M.~Zaro,
%``The automated computation of tree-level and next-to-leading order differential cross sections, and their matching to parton shower simulations,''
JHEP \textbf{07}, 079 (2014)
doi:10.1007/JHEP07(2014)079
[arXiv:1405.0301 [hep-ph]].
%9074 citations counted in INSPIRE as of 28 Dec 2024

%\cite{Pumplin:2002vw}
\bibitem{Pumplin:2002vw}
J.~Pumplin, D.~R.~Stump, J.~Huston, H.~L.~Lai, P.~M.~Nadolsky and W.~K.~Tung,
%``New generation of parton distributions with uncertainties from global QCD analysis,''
JHEP \textbf{07}, 012 (2002)
doi:10.1088/1126-6708/2002/07/012
[arXiv:hep-ph/0201195 [hep-ph]].
%7067 citations counted in INSPIRE as of 24 Dec 2024

%\cite{Bierlich:2022pfr}
\bibitem{Bierlich:2022pfr}
C.~Bierlich, S.~Chakraborty, N.~Desai, L.~Gellersen, I.~Helenius, P.~Ilten, L.~L\"onnblad, S.~Mrenna, S.~Prestel and C.~T.~Preuss, \textit{et al.}
%``A comprehensive guide to the physics and usage of PYTHIA 8.3,''
SciPost Phys. Codeb. \textbf{2022}, 8 (2022)
doi:10.21468/SciPostPhysCodeb.8
[arXiv:2203.11601 [hep-ph]].
%729 citations counted in INSPIRE as of 27 Dec 2024

%\cite{Mangano:2006rw}
\bibitem{Mangano:2006rw}
M.~L.~Mangano, M.~Moretti, F.~Piccinini and M.~Treccani,
%``Matching matrix elements and shower evolution for top-quark production in hadronic collisions,''
JHEP \textbf{01}, 013 (2007)
doi:10.1088/1126-6708/2007/01/013
[arXiv:hep-ph/0611129 [hep-ph]].
%915 citations counted in INSPIRE as of 18 Dec 2024

%\cite{Hoeche:2005vzu}
\bibitem{Hoeche:2005vzu}
S.~Hoeche, F.~Krauss, N.~Lavesson, L.~Lonnblad, M.~Mangano, A.~Schalicke and S.~Schumann,
%``Matching parton showers and matrix elements,''
doi:10.5170/CERN-2005-014.288
[arXiv:hep-ph/0602031 [hep-ph]].
%406 citations counted in INSPIRE as of 02 Dec 2024

%\cite{deFavereau:2013fsa}
\bibitem{deFavereau:2013fsa}
J.~de Favereau \textit{et al.} [DELPHES 3],
%``DELPHES 3, A modular framework for fast simulation of a generic collider experiment,''
JHEP \textbf{02}, 057 (2014)
doi:10.1007/JHEP02(2014)057
[arXiv:1307.6346 [hep-ex]].
%3106 citations counted in INSPIRE as of 28 Dec 2024

%\cite{Selvaggi:2014mya}
\bibitem{Selvaggi:2014mya}
M.~Selvaggi,
%``DELPHES 3: A modular framework for fast-simulation of generic collider experiments,''
J. Phys. Conf. Ser. \textbf{523}, 012033 (2014)
doi:10.1088/1742-6596/523/1/012033
%118 citations counted in INSPIRE as of 06 Nov 2024

%\cite{Mertens:2015kba}
\bibitem{Mertens:2015kba}
A.~Mertens,
%``New features in Delphes 3,''
J. Phys. Conf. Ser. \textbf{608}, no.1, 012045 (2015)
doi:10.1088/1742-6596/608/1/012045
%86 citations counted in INSPIRE as of 20 Nov 2024

%\cite{Cacciari:2008gp}
\bibitem{Cacciari:2008gp}
M.~Cacciari, G.~P.~Salam and G.~Soyez,
%``The anti-$k_t$ jet clustering algorithm,''
JHEP \textbf{04}, 063 (2008)
doi:10.1088/1126-6708/2008/04/063
[arXiv:0802.1189 [hep-ph]].
%10571 citations counted in INSPIRE as of 29 Dec 2024

%\cite{Cacciari:2011ma}
\bibitem{Cacciari:2011ma}
M.~Cacciari, G.~P.~Salam and G.~Soyez,
%``FastJet User Manual,''
Eur. Phys. J. C \textbf{72}, 1896 (2012)
doi:10.1140/epjc/s10052-012-1896-2
[arXiv:1111.6097 [hep-ph]].
%6275 citations counted in INSPIRE as of 29 Dec 2024

%\cite{DARWIN:2016hyl}
\bibitem{DARWIN:2016hyl}
J.~Aalbers \textit{et al.} [DARWIN],
%``DARWIN: towards the ultimate dark matter detector,''
JCAP \textbf{11}, 017 (2016)
doi:10.1088/1475-7516/2016/11/017
[arXiv:1606.07001 [astro-ph.IM]].
%778 citations counted in INSPIRE as of 24 Dec 2024

%\cite{Esteban:2024eli}
\bibitem{Esteban:2024eli}
I.~Esteban, M.~C.~Gonzalez-Garcia, M.~Maltoni, I.~Martinez-Soler, J.~P.~Pinheiro and T.~Schwetz,
%``NuFit-6.0: Updated global analysis of three-flavor neutrino oscillations,''
[arXiv:2410.05380 [hep-ph]].
%34 citations counted in INSPIRE as of 25 Dec 2024

%\cite{Baglio:2015wcg}
\bibitem{Baglio:2015wcg}
J.~Baglio, A.~Djouadi and J.~Quevillon,
%``Prospects for Higgs physics at energies up to 100 TeV,''
Rept. Prog. Phys. \textbf{79}, no.11, 116201 (2016)
doi:10.1088/0034-4885/79/11/116201
[arXiv:1511.07853 [hep-ph]].
%63 citations counted in INSPIRE as of 16 Dec 2024

%\cite{CMS:2024qxz}
\bibitem{CMS:2024qxz}
A.~Hayrapetyan \textit{et al.} [CMS],
%``Search for long-lived particles decaying to final states with a pair of muons in proton-proton collisions at $ \sqrt{s} $ = 13.6 TeV,''
JHEP \textbf{05}, 047 (2024)
doi:10.1007/JHEP05(2024)047
[arXiv:2402.14491 [hep-ex]].
%10 citations counted in INSPIRE as of 07 Apr 2025

%\cite{CMS:2021juv}
\bibitem{CMS:2021juv}
A.~Tumasyan \textit{et al.} [CMS],
%``Search for Long-Lived Particles Decaying in the CMS End Cap Muon Detectors in Proton-Proton Collisions at $\sqrt s$ =13\,\,TeV,''
Phys. Rev. Lett. \textbf{127}, no.26, 261804 (2021)
doi:10.1103/PhysRevLett.127.261804
[arXiv:2107.04838 [hep-ex]].
%72 citations counted in INSPIRE as of 07 Apr 2025

%\cite{CMS:2024hik}
\bibitem{CMS:2024hik}
A.~Hayrapetyan \textit{et al.} [CMS],
%``Search for long-lived heavy neutral leptons in proton-proton collision events with a lepton-jet pair associated with a secondary vertex at $ \sqrt{s} $ = 13 TeV,''
JHEP \textbf{02}, 036 (2025)
doi:10.1007/JHEP02(2025)036
[arXiv:2407.10717 [hep-ex]].
%12 citations counted in INSPIRE as of 07 Apr 2025


\end{thebibliography}
\end{document}